\documentclass[fleqn,usenatbib]{mnras}

\usepackage{newtxtext,newtxmath}

\usepackage[T1]{fontenc}

\DeclareRobustCommand{\VAN}[3]{#2}
\let\VANthebibliography\thebibliography
\def\thebibliography{\DeclareRobustCommand{\VAN}[3]{##3}\VANthebibliography}

\usepackage{graphicx}	
\usepackage{amsmath}	
\usepackage{xspace}
\usepackage{orcidlink}
\usepackage{comment}
\usepackage{dcolumn}
\usepackage{multirow}
\usepackage{siunitx}
\usepackage{xcolor}
\usepackage{tablefootnote}
\usepackage[flushleft]{threeparttable}

\newcommand{\pcm}{\,cm$^{-3}$\xspace}	
\newcommand\gsim{~\lower.5ex\hbox{$\buildrel > \over \sim$}~}
\newcommand\lsim{~\lower.5ex\hbox{$\buildrel < \over \sim$}~}
\newcommand{\lsun}{L$_\odot$\xspace}

\newcommand{\vlsr}{v$_{\rm{lsr}}$\xspace}

\newcommand{\micr}{$\mu$m\xspace}
\newcommand{\kms}{km\,s$^{-1}$\xspace}

\newcommand{\hii}{H\textsc{ii}\xspace}
\newcommand{\uchii}{UC\,H\,{\sc ii}\xspace}

\newcommand{\simm}{$\sim$}

\newcommand{\higal}{Hi-GAL\xspace}

\newcolumntype{d}[1]{D{.}{\cdot}{#1}}
\definecolor{mygray}{gray}{0.6}
\newcolumntype{.}{D{.}{.}{-1}}

\title[The HCN-IR Luminosity Relation]{ATLASGAL: The HCN-IR Star Formation Relation for the Milky Way\thanks{The full version of Table\,\ref{tbl:clump_properties} is only available in electronic form at the CDS via anonymous ftp to cdsarc.u-strasbg.fr (130.79.125.5) or via http://cdsweb.u-strasbg.fr/cgi-bin/qcat?J/MNRAS/.}}

\author[I.\,I.\,Grozdanova et al.]{I.\,I.\,Grozdanova\orcidlink{0009-0006-6327-0595},$^{1}$\thanks{E-mail: i.grozdanova@kent.ac.uk}
J.\,S.\,Urquhart\orcidlink{0000-0002-1605-8050},$^{1}$\thanks{E-mail: j.s.urquhart@kent.ac.uk}
W.-J.\,Kim \orcidlink{0000-0003-0364-6715},$^{2}$
F.\,Wyrowski,$^{2}$
T.\,Csengeri,$^{3}$
A.\,Karska \orcidlink{0000-0001-8913-925X},$^{4,2}$
\newauthor
A.\,Giannetti,$^{5}$
D.\,Colombo\orcidlink{0000-0001-6498-2945},$^{2,6}$
D.\,Eden \orcidlink{0000-0002-5881-3229},$^{7}$
M.\,A.\,Thompson,$^{8}$
T.\,J.\,T.\,Moore$^{9}$
\\
$^{1}$ Centre for Astrophysics and Planetary Science, University of Kent, Canterbury, CT2 7NH, UK \\
$^{2}$  Max-Planck-Institut f\"ur Radioastronomie (MPIfR), Auf dem H\"ugel 69, 53121 Bonn, Germany \\
$^{3}$ Laboratoire d’astrophysique de Bordeaux, Univ. Bordeaux, CNRS, B18N, allée Geoffroy Saint-Hilaire, F-33615 Pessac, France\\
$^{4}$ Institute of Advanced Studies, Nicolaus Copernicus University in Toruń, Wileńska 4, 87-100 Toruń, Poland\\
$^{5}$ INAF - Istituto di Radioastronomia, Via P. Gobetti 101, I-40129 Bologna, Italy\\
$^{6}$ Argelander-Institut f\"ur Astronomie, University of Bonn, Auf dem H\"ugel 71, 53121 Bonn, Germany\\
$^{7}$ Department of Physics, University of Bath, Claverton Down, Bath BA2\,7AY, UK\\
$^{8}$ School of Physics and Astronomy, University of Leeds, Leeds LS2 9JT, UK\\
$^{9}$Astrophysics Research Institute, Liverpool John Moores University, IC2, Liverpool Science Park, 146 Brownlow Hill, Liverpool L3 5RF, UK
}

\date{Accepted XXX. Received YYY; in original form ZZZ}

\pubyear{2026}

\begin{document}
\label{firstpage}
\pagerange{\pageref{firstpage}--\pageref{lastpage}}
\maketitle

\begin{abstract}
    We examine the relation between star formation and dense gas, traced by HCN\,(1–0), using a sample of 343 Galactic clumps spanning the full range of evolutionary stages representative of high-mass and cluster formation, including quiescent, protostellar, young stellar object (YSO), and \hii regions. Earlier Galactic studies, based primarily on luminous \hii regions, reported a linear correlation between infrared (IR) and HCN luminosities, consistent with extragalactic surveys, and interpreted it as a universal star formation law. Our larger evolutionary sample yields a significantly steeper slope of $1.80 \pm 0.08$, inconsistent with extragalactic trends. However, we find that each of the star-forming evolutionary stages individually has a shallower slope (\simm1.2) than that of the full sample, with quiescent clumps showing no correlation. We estimate the Milky Way’s total IR and HCN luminosities from dense clumps by integrating the bolometric luminosities and clump masses of \higal sources. Both fall significantly below extragalactic values, demonstrating that clumps alone cannot account for the IR–HCN relation. Only \simm5--25\,per\,cent of Galactic HCN luminosity arises from dense clumps, implying that most extragalactic HCN traces gas that is not involved in star formation. Additionally, dense clumps contribute only \simm2–3\,per\,cent of the Milky Way’s IR luminosity, which is dominated instead by extended \hii regions and PDRs tracing longer star formation timescales. As a result, the observed linear relation in nearby galaxies reflects a nearly constant star formation efficiency averaged over large areas and long timescales, rather than a direct link between SFR and dense gas mass.
\end{abstract}


\section{Introduction} \label{Introduction}

Despite massive stars (O- and B-type stars; M$_\star$ > 8 M$_{\odot}$) being extremely rare, they contribute the majority of a galaxy's luminosity \citep{McKee&Ostriker2007}. Throughout their life, high-mass stars enrich the interstellar medium with metals and can even trigger star formation throughout a galaxy \citep{Thompson2012} driving its evolution \citep{Larson1998}. Stars are born in cold molecular clouds ($\sim$10\,K) that predominantly consist of molecular hydrogen (H$_2$). The distribution and density of H$_2$ can be estimated using tracer molecules that are excited through collisions with the molecular hydrogen in the gas. The most commonly used molecular tracers in both Galactic and extra-galactic astronomy are the rotational transitions of carbon monoxide ($^{12}$CO), which is the second most abundant molecule in the interstellar medium (ISM) and is easily excited at low densities (see \citealt{Bolatto2013} for a detailed review on the use of CO as a gas tracer).

Star formation in galaxies has been shown to be related to the gas volumetric density in the form of the Schmidt power law \citep{Schmidt1959}. This relationship has been extended to several orders of magnitude by examining different types of galaxies. Using a sample of 61 normal disc galaxies and 36 infrared-selected starburst galaxies, \citet{Kennicutt1998} found that their sample is well described by a power-law with an index of $N=1.4\pm0.15$; this is generally referred to as the Kennicutt-Schmidt relation. This relationship can be interpreted as a direct relationship between the star formation rate (SFR), derived from H$\alpha$ emission-line flux, and the amount of molecular hydrogen gas in the galaxy. The SFR quantifies the rate at which gas is converted into stars, and while H$\alpha$ traces the ionising radiation of young massive stars directly, it is strongly affected by dust extinction. Infrared (IR) luminosity, however, arises from dust that is directly heated by the radiation of embedded protostellar objects and re-emitted, making it a more reliable tracer of recent star formation, particularly in the dust-obscured regions where massive stars form \citep{Kennicut+Evans2012}. The star formation efficiency (SFE$\equiv$SFR/M$_{\rm gas}$), the amount of gas converted into stars, is therefore closely tied to the Kennicutt-Schmidt relation, which describes how the star-forming process scales with gas surface density \citep{Kennicutt1998}.

In practice, the gas surface density is typically measured using $^{12}$CO\,(1--0) emission as a proxy for the total molecular gas reservoir. However, $^{12}$CO\,(1--0) traces gas with number densities as low as $n(\text{H}_2)=300$\pcm, far below typical dense-gas thresholds for star formation ($n(\text{H}_2) \gtrsim 10^4$\pcm; e.g., \citealt{Lada2010}). Therefore, not all of the gas traced by the CO in these studies participates in the formation of stars. The fraction of total gas mass above the density threshold typically required for star formation, commonly referred to as the dense-gas fraction ($f_{\rm dense}\equiv M_{\rm dense}/M_{\rm gas}$), has been found to correlate more tightly with SFR than total molecular gas mass alone \citep{Gao+Solomon2004}, highlighting the importance of tracers sensitive specifically to this dense component. To better trace the dense gas, where stars are known to form, emission from molecules such as CS and HCN that are considered to trace significantly higher gas densities are employed \citep{Shirley2015}. HCN\,(1--0) and CS\,(2--1) (hereafter referred to as CS and HCN unless otherwise specified), with critical densities of $n_{\text{crit}}(\text{H}_2) \gtrsim 10^5$\pcm at gas temperatures below 40\,K \citep{Shirley2015}, are commonly assumed to be sensitive to dense molecular gas and are easily observed both within the Milky Way and in external galaxies.

As discussed above, IR luminosity is widely used as a tracer of star formation, and high-critical-density molecules such as HCN are often assumed to trace the gas most directly available for star formation. This has motivated numerous studies of the HCN–IR luminosity relationship as a probe of the connection between dense gas and star formation. \citet{Gao+Solomon2004} used HCN\,(1--0) to study the dense gas associated with a sample of 65 nearby galaxies with strong $^{12}$CO and/or infrared (IR) emission. They found a correlation between infrared luminosity ($L_{\text{IR}}$) and the HCN molecular line luminosity ($L_{\text{HCN}}$) with a slope of $1.00\pm0.05$, which extends over 3 orders of magnitude. They show this relationship has less scatter than the correlation between $L_{\text{IR}}$ and $L_{\text{$^{12}$CO}}$, which has a slope of $1.25\pm0.08$ and steepens at higher IR luminosities. The results of \citet{Gao+Solomon2004} demonstrate that the Kennicutt-Schmidt relation between $^{12}$CO and SFR is in fact likely due to the tight relationship between total gas (as traced by CO) and dense gas (traced by HCN), and not because of a direct relationship between the CO emission and the star formation rate. This led the authors to argue that HCN luminosity can be used as a more reliable star formation indicator.

\begin{figure}
    \centering
    \includegraphics[width = 0.45\textwidth]{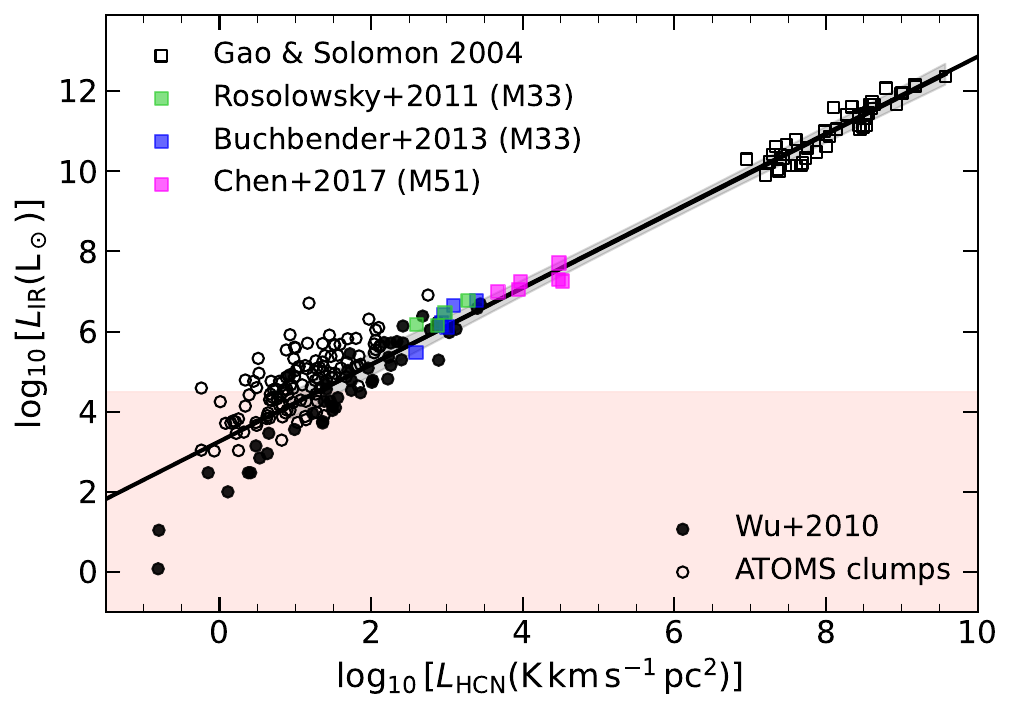}
    \caption{Relationship between IR and HCN luminosities as traced by the  \citet{Gao+Solomon2004} and \citet{Wu2010} samples. Clumps below $L_{\rm IR}=10^{4.5}$\,\lsun\ (marked by a rosy region) are not used in the fit, to be consistent with \citet{Wu2005, Wu2010}. The black fitted line has a slope of $0.96\pm0.01$ and the shaded region around it represents the $1\sigma$ uncertainty in the fit. Additionally, some Galactic \citep[ATOMS survey;][]{ATOMS2} and extra-galactic studies \citep{Rosolowsky2011, Buchbender2013, Chen2017} have been added for comparison.}
    \label{fig:Wu_slopes}
\end{figure}

Their work was followed by a study of 47 Galactic star-forming clumps by \citet{Wu2005, Wu2010}, selected from a larger sample of sources associated with water masers. Using a least-squares fit to the clumps with IR luminosities above $L_{\text{IR}}>10^{4.5}$\,L$_\odot$ they determined a linear relationship (with a slope of $1.02\pm 0.06$) that is consistent with the \citet{Gao+Solomon2004} slope and effectively extends the correlation to over eight orders of magnitude in HCN and IR luminosities. They conclude that this is indicative of a constant star-formation rate per unit of dense gas mass for all scales from dense clumps to entire galaxies, extending the relationship between Galactic and extragalactic clouds.

In Figure\,\ref{fig:Wu_slopes}, we reproduce the correlation plot presented in \citet{Wu2005}, excluding clumps with IR luminosities less than 10$^{4.5}$\,\lsun from the fit as they did. There are slight differences in the HCN luminosities given in \citet{Wu2005} and \citet{Wu2010} and so we have refitted the slope using the values given in \citet{Wu2010} and \citet{Gao+Solomon2004} data. This gives a slope of $0.96\pm0.01$ (black solid line on Figure\,\ref{fig:Wu_slopes}), which is shallower than reported by \citet{Wu2005} but still broadly linear. However, we note that only fitting the data above the 10$^{4.5}$\,\lsun threshold biases the sample to bright \uchii regions and so is unlikely to be representative of the general population of Galactic clumps. Although the brightest star-forming regions dominate the infrared luminosity, they represent only a small fraction of the overall clump population. The bulk of the molecular mass instead resides in the far more numerous clumps at earlier evolutionary stages \citep{Elia2021}.

The work by \citet{Wu2010} was followed up with a larger sample by \citet{Stephens2016} using data from the Millimetre Astronomy Legacy Team 90-\unit{\giga\hertz} catalogue \citep[MALT90;][]{Jackson2013, Rathborne2016}. This survey mapped 3246 star-forming clumps as previously identified by the APEX Telescope Large Area Survey of the Galaxy \citep[ATLASGAL; ][]{ATLASGAL2009}. \citet{Stephens2016} successfully matched 405 MALT90 clumps with an infrared source found in the IRAS Point Source v2.1 catalogue \citep{IRAS1988, IRAS2019}. Of these, \simm160 were detected in HCN. The HCN luminosity was measured directly from the MALT90 maps while the infrared luminosity was estimated following the method outlined by \citet{Sanders&Mirabel1996} using four IRAS bands (the same method was used by \citealt{Gao+Solomon2004} and \citealt{Wu2010}).

Fitting their whole sample of Galactic clumps, \citet{Stephens2016} obtained a slope of $1.21\pm0.07$, which is steeper than the \citet{Gao+Solomon2004} and \citet{Wu2010} relation, but their sample includes a wider range of evolutionary stages and the slope is less well constrained. \citet{Stephens2016} also noted a trend based on the evolutionary stage of the clumps, with \hii regions tending to lie above the least squares fit to their sample, while protostellar clumps tend to be below the fitted slope.

More recently, the `ALMA Three-millimeter Observations of Massive Star-forming regions' (ATOMS) survey \citep{ATOMS1} has investigated 119 clumps with IRAS colours characteristic of \mbox{\uchii\,regions} \citep{ATOMS2} with angular and spectral resolution of \simm13.5\,arcsec and \simm0.2\,\kms respectively. They produce integrated intensity maps of the HCN\,(1--0) emission and find that ATOMS clumps follow a linear IR-HCN luminosity relationship with a slope of $0.99\pm0.08$. The similarity in the ATOMS and \citet{Wu2010} slopes is not altogether surprising given that both samples are selected using the same IRAS colour criterion. Although the slopes are similar, the ATOMS clumps sit above the Gao-Solomon relation and have $L_{\rm IR}/L_{\rm HCN}$ ratios that are \simm5 times larger (see distribution of open circles in Figure\,\ref{fig:Wu_slopes}). However, \citet{ATOMS2} suspect that this is due to the nature of the interferometric observations, which filter out larger angular scales, resulting in lower HCN fluxes than would be detected in single dish observations.

While the resolution we can achieve in our own Galaxy is unmatched by extragalactic studies, there are a few nearby galaxies that are close enough to resolve individual star-forming clouds. In Figure\,\ref{fig:Wu_slopes}, we also included results from high-resolution observations of M51 \citep{Chen2017} and M33 \citep{Rosolowsky2011,Buchbender2013}. With spatial resolution of $\sim100-150$\,pc, these studies focus on giant molecular cloud (GMC) scales and include HCN emission from both dense and sub-thermally excited (i.e.\,gas below the critical density of HCN) gas and infrared emission from both embedded clumps and extended \hii\ regions. These extragalactic GMCs generally follow the relationship found for whole galaxies, albeit with some scatter, and give weight to the idea of a universal star-formation relation. A more complete compilation of Galactic and extra-galactic observations can be found in \citet{Neumann2025}.

To better understand the physical mechanism behind the Gao-Solomon relation, \citet{Wu2005}, \citet{Stephens2016} and \citet{ATOMS2} have extended the relationship down to the scale of individual star-forming clumps. However, their focus on clumps with counterparts in the IRAS catalogue inherently biases the sample toward the most luminous clumps in the Galaxy, many of which are likely to be associated with embedded \hii\ regions. In this work, we aim to explore the Gao–Solomon relation in greater detail by expanding the sample size and including clumps at earlier evolutionary stages. If the extra-galactic relationship indeed demonstrates the link between SFR and dense molecular material, then any star-forming clump should follow the same trend, and all of the IR and HCN emission of galaxies must originate from star-forming clumps. We analyse HCN emission from $\sim$400 ATLASGAL clumps observed with the IRAM-30m telescope (\citealt{Csengeri2016, Kim2020}). We combine this sample with the MALT90 dataset analysed by \citet{Stephens2016} and investigate the relationship between dense gas and star formation rate with a large and representative sample of dense Galactic clumps that includes the full range of evolutionary stages.

The structure of this paper is as follows: in Section\,\ref{observations}, we discuss the sample selection, observations, and data reduction. In Section\,\ref{results}, we present the analysis of the spectral line profiles, calculate the molecular line luminosity, and determine a conversion between $L_{\text{IR}}$ and $L_{\text{bol}}$, as well as $L_{\text{HCN}}$ and $M_{\text{dense}}$ that is required to compare our results with those in the literature. In Section\,\ref{SF relation}, we investigate the relationship between HCN and IR luminosity and compare our sample to previous studies of Galactic clumps. An analysis of the results in the context of extragalactic observations is provided in Section\,\ref{sect:milkyway}. A summary of the findings is given in Section\,\ref{Conclusion}.

\begin{figure*}
    \centering
    \includegraphics[width = 0.45\textwidth]{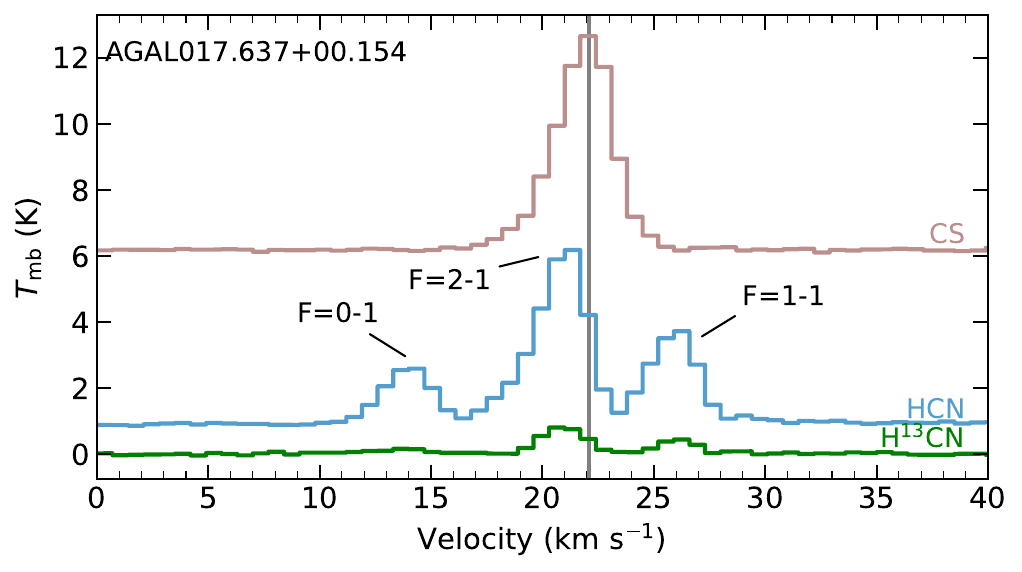}
    \includegraphics[width = 0.45\textwidth]{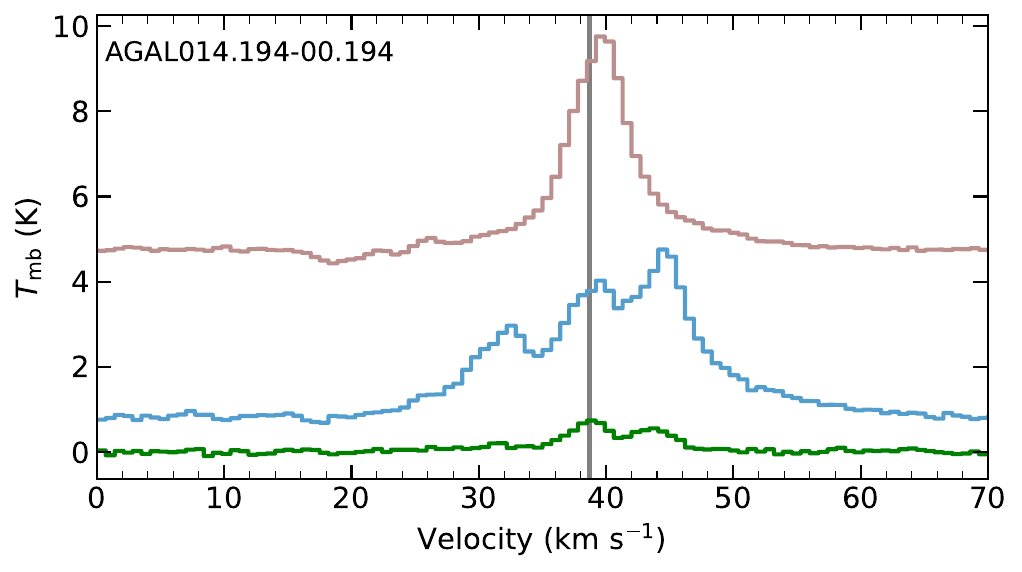}\\

    \includegraphics[width = 0.45\textwidth]{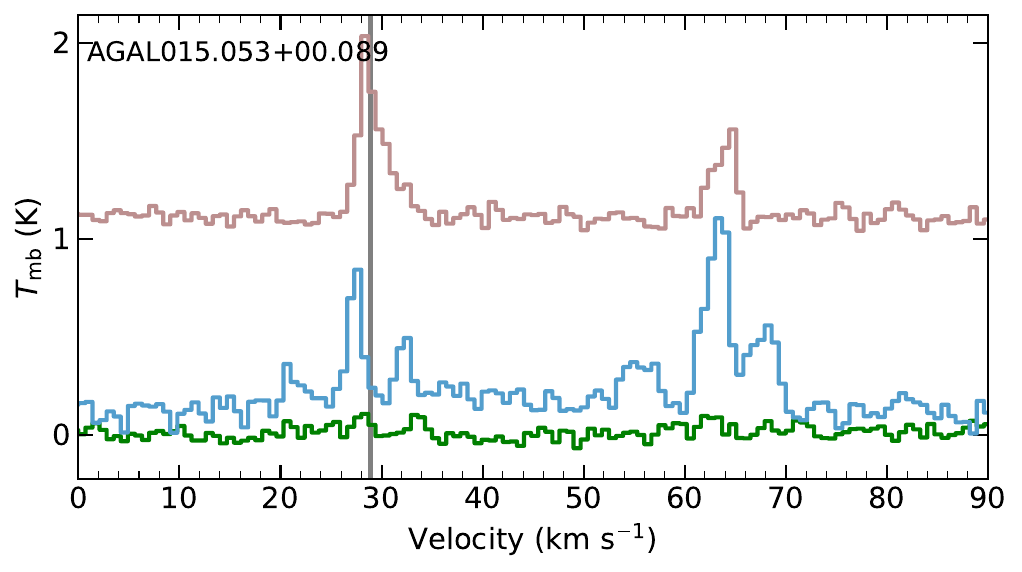}
    \includegraphics[width = 0.45\textwidth]{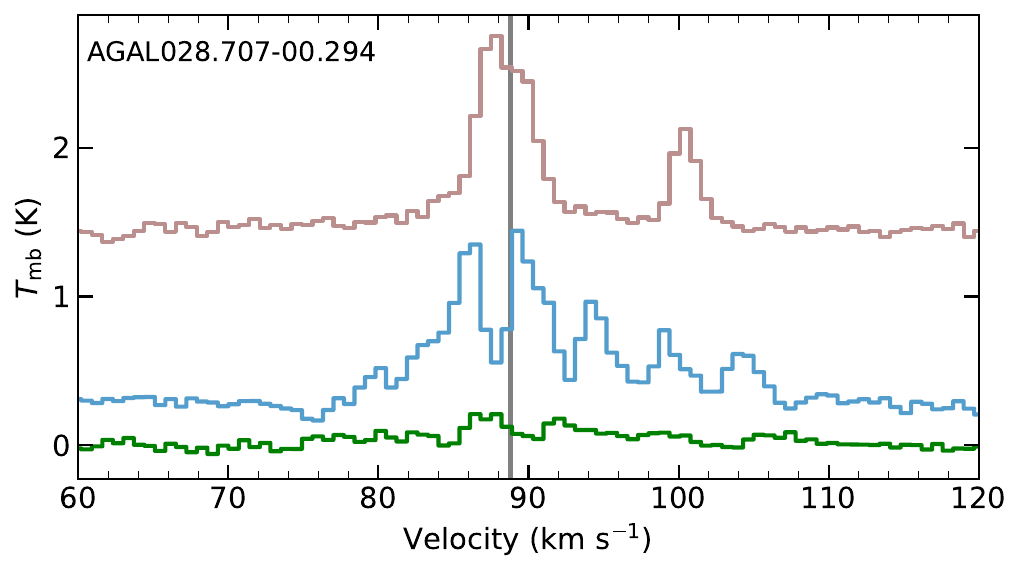}\\

    \includegraphics[width = 0.45\textwidth]{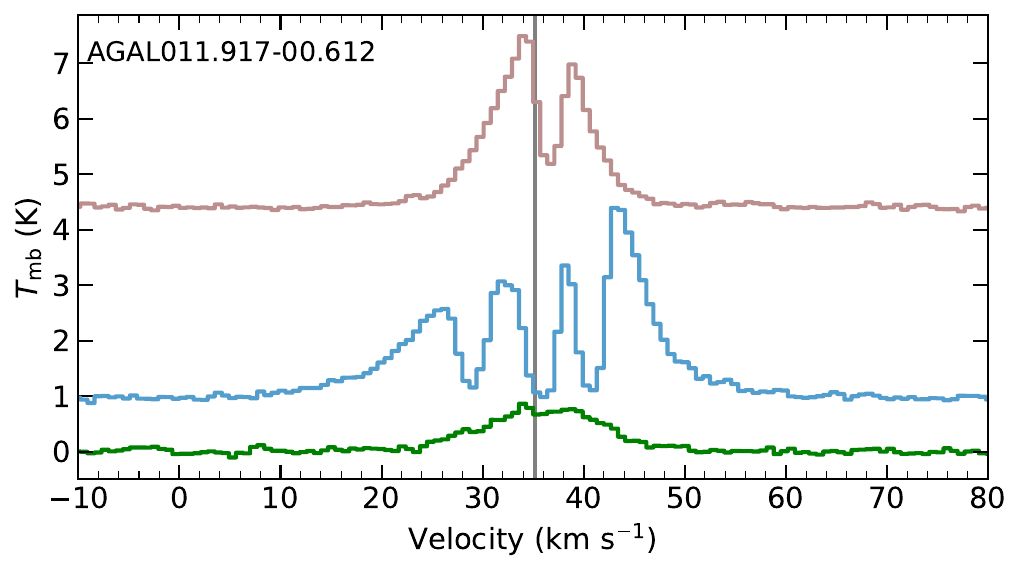}
    \includegraphics[width = 0.45\textwidth]{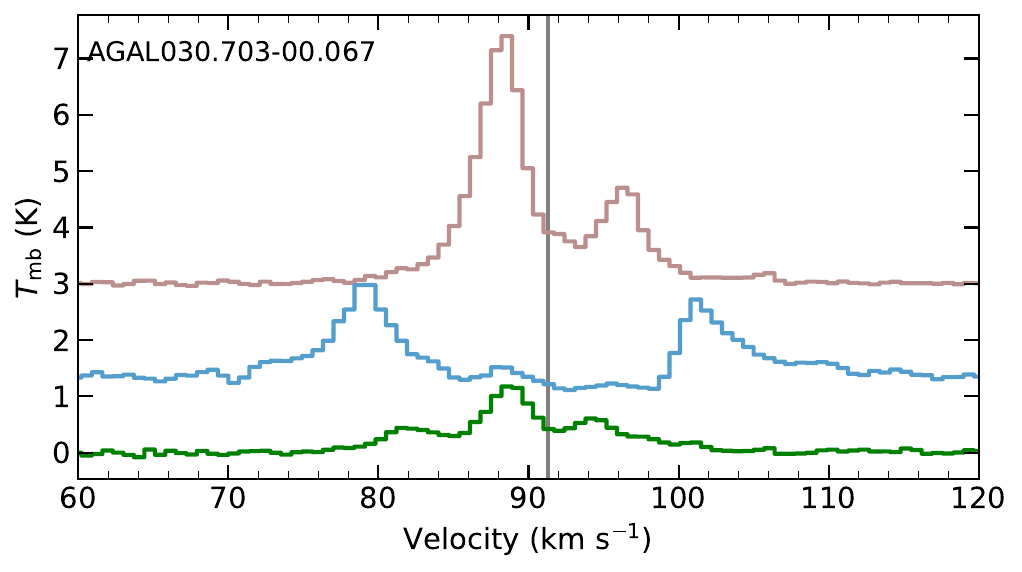}\\

    \includegraphics[width = 0.45\textwidth]{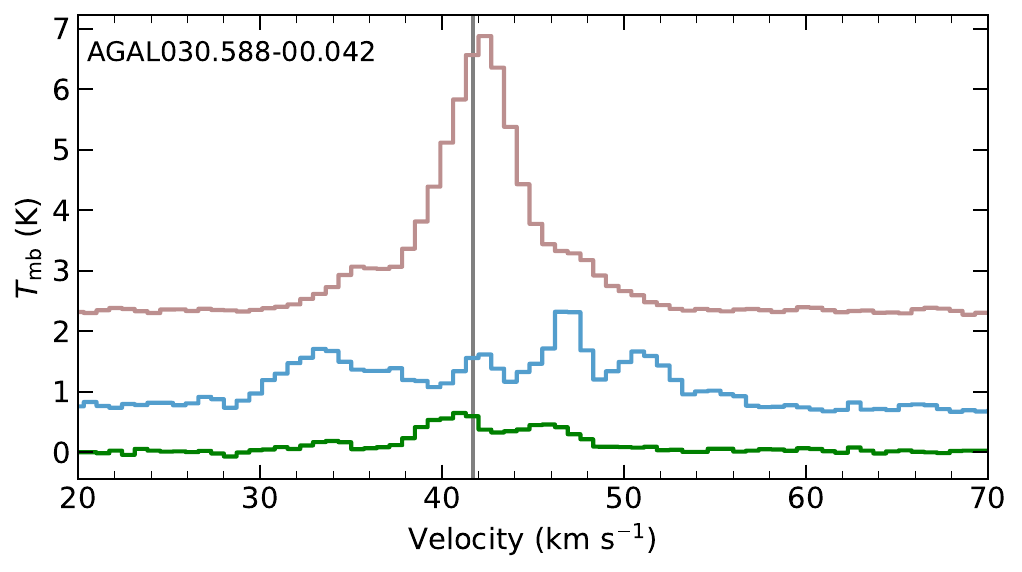}
    \includegraphics[width = 0.45\textwidth]{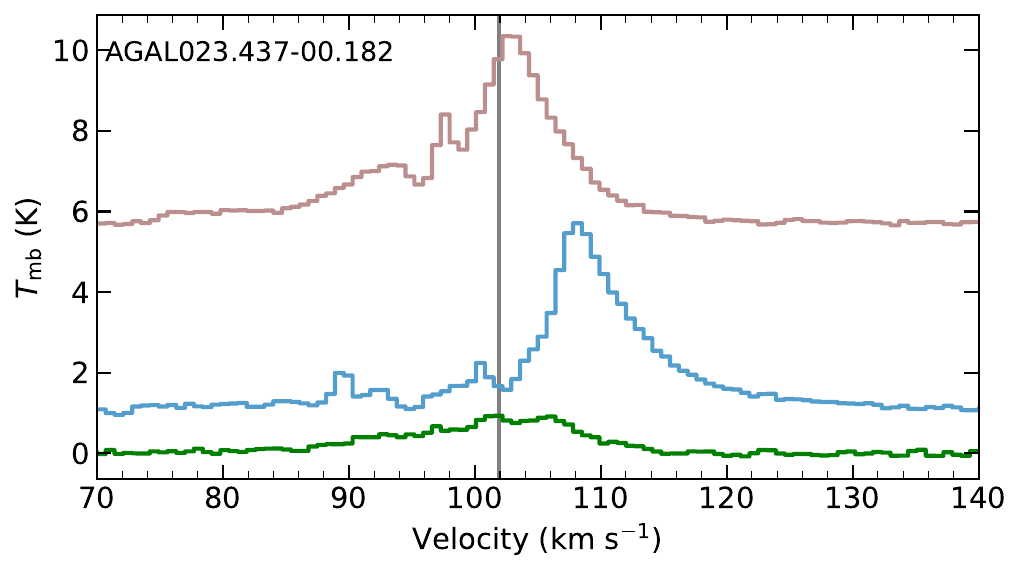}\\

    \caption{Example spectra (H$^{13}$CN (\textit{green}), HCN (\textit{blue}), CS (\textit{red})) of several clumps. A vertical line indicates the previously determined clump velocity \citep{Urquhart2022}. \textit{Top panels:} An ideal spectral profile where the HCN line has 3 clearly separated fine-line features (\textit{left}) and an example of anomalous HCN profile (\textit{right}); \textit{Second row:} Both panels show sources with emission at two distinct velocities. The left panel shows a case in which the two clumps are clearly separated for both molecular gas tracers. In this case the stronger CS line matches the velocity in the ATLASGAL velocity but the stronger HCN emission is actually associated with a weaker CS peak. The right-hand-side panel demonstrates that due to the complex structure of HCN emission, two clumps that appear distinct in CS are not necessarily so when only examining the HCN profile, as a result of blending of the hyperfine components; \textit{Third row:} Two examples of self-absorption features in the HCN emission; \textit{Lower panels:} Two examples of HCN profiles we would classify as uncertain.}
    \label{fig:V_disc_examples}
\end{figure*}

\section{Observations and Data Processing } \label{observations}

\subsection{Observations of an ATLASGAL sample}

ATLASGAL \citep{ATLASGAL2009, Csengeri2014} provides an unbiased survey
of the dense dust emission across the inner Galactic plane ($300$\degr$<\ell<60$\degr\ and $|b|<1.5$\degr) using 870-\micr  submillimetre continuum observations. The compact-source catalogue (CSC) produced from the survey maps contains $\sim10\,000$ dense clumps \citep[][]{Contreras2013, Urquhart2014}. This catalogue covers the full range of evolutionary stages from quiescent clumps to clumps hosting \hii regions \citep{Urquhart2013, Urquhart2014_atlas,Urquhart2018, Urquhart2022}.

A sample of 430 ATLASGAL clumps was selected for follow-up observations with the IRAM 30-m telescope (project codes: 181-10 and 049-11; reported in \citealt{Csengeri2016, Kim2020}) using the EMIR receiver \citep{EMIR}. These clumps, located in the 1$^{\rm st}$ Quadrant ($5\degr <\ell < 60\degr$), were chosen as part of an unbiased spectral line survey to allow the investigation of dynamical properties and chemistry as a function of clump evolution (i.e., quiescent, protostellar, young stellar object (YSO) and \hii region clumps). The spectroscopic line data were obtained using the On-Off position-switching (PSw) mode to produce calibrated antenna temperatures ($T_{\mathrm{A}}^*$) in Kelvin. The observations cover a wide range of frequencies ($\sim84\,\unit{\giga\hertz} - 115\,\unit{\giga\hertz}$) with a beam size of 29\,arcsec. All spectral line data were resampled to a uniform velocity resolution of 0.7 km s$^{-1}$ before extracting individual spectra, using the Continuum and Line Analysis Single-dish Software (CLASS\footnote{\url{https://www.iram.fr/IRAMFR/GILDAS/doc/html/class-html/class.html}}) within the Grenoble Image and Line Data Analysis Software (GILDAS) package \citep{Pety2005_gildas}. The frequency range includes a large number of molecular transitions (see \citealt{Csengeri2016} and \citealt{Kim2020} for a detailed description of the observations and analysis of several molecular transitions), including many high-density tracers; however, studies of other galaxies are currently lacking many of these. In this work, therefore, we concentrate on the HCN\,(1--0) transition to allow detailed comparisons with previous Galactic and extra-galactic studies (\citealt{Gao+Solomon2004, Wu2005, Stephens2016}).

\subsection{HCN detections and line profile analysis} \label{data_red}

Of the 430 sources observed for this project, 426 were observed in the setup that covers the HCN transition (for more observational details, see \citealt{Csengeri2016} and \citealt{Kim2020}). The positions of the observations have then been matched with the ATLASGAL catalogue using a 30-arcsec search radius resulting in 390 unique matches.\footnote{The target list was compiled using a preliminary version of the ATLASGAL catalogue and some of the low surface brightness clumps did not make it into the final catalogue.}

A first-order polynomial baseline has been fitted to the HCN spectrum of each clump and subtracted to account for any deviation of the mean noise from zero. A forward efficiency of 0.95 and a beam efficiency of 0.81 have been used to convert between $T_\text{A}^\star$ and $T_{\text{mb}}$ following \citet{Csengeri2016}. An emission line has been defined as a detection if there are at least three consecutive channels above the 3-$\sigma$ noise threshold (resulting in a minimum line width of 2.1\,\kms), where the noise is determined from the standard deviation of the emission-free channels and has a mean value for our sample of \simm0.04\,K. This applied detection threshold resulted in HCN emission being detected towards 381 clumps, with 9 non-detections.

In Figure\,\ref{fig:V_disc_examples}, we show examples of HCN spectra detected towards ATLASGAL sources and include the H$^{13}$CN (1--0) and CS (2--1) transitions, which are simultaneously obtained from the same observational data for HCN, as a comparison. The top left panel shows a typical profile where the HCN rotational levels show hyperfine structure (HFS); this consists of three distinct emission components that are a result of the presence of the nitrogen atom. The three components have relative intensities of 1:5:3 and a separation of $-7.1$\,\unit{\kilo\meter\per\second} and 4.9\,\unit{\kilo\meter\per\second} for $F=0-1$ ($\nu=88633.9357$\,MHz) and $F=1-1$ ($\nu=88630.4156$\,MHz) from the central $F=2-1$ ($\nu=88631.8475$\,MHz) line, respectively \citep{Goicoechea2022}.

\begin{figure*}
    \centering
    \includegraphics[width = 0.45\textwidth]{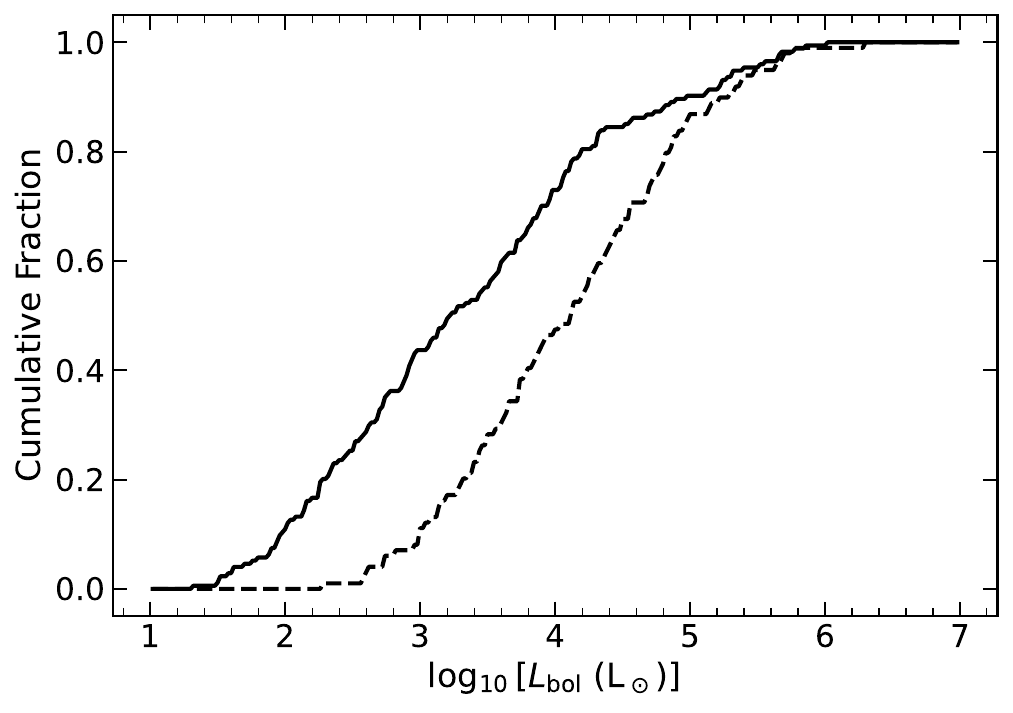}
    \includegraphics[width = 0.45\textwidth]{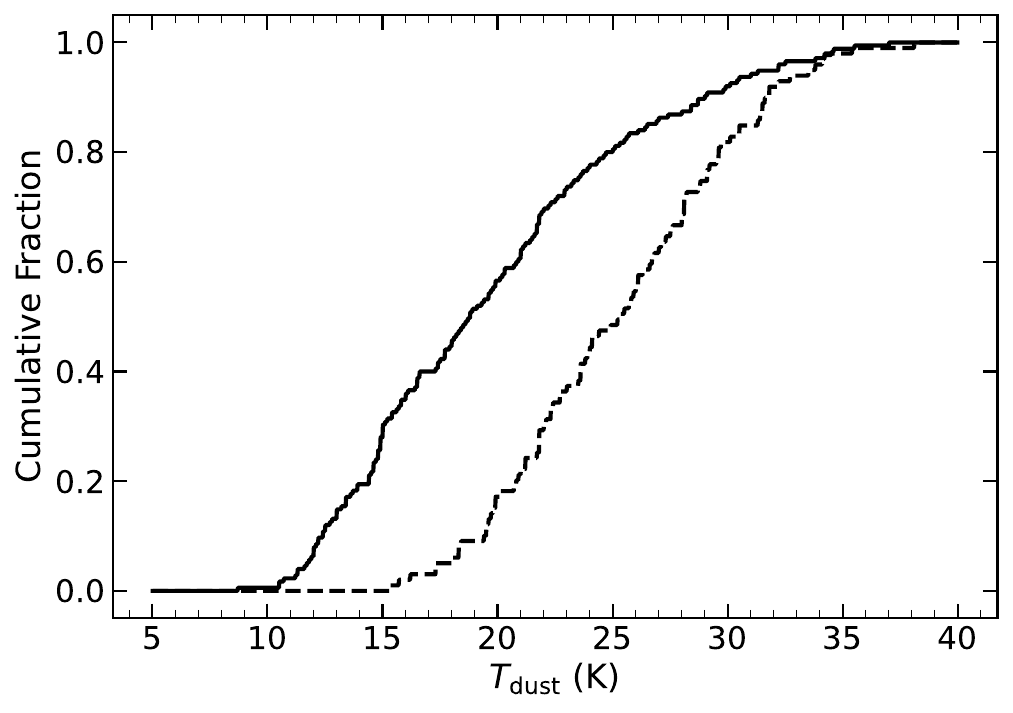}\\

    \includegraphics[width = 0.45\textwidth]{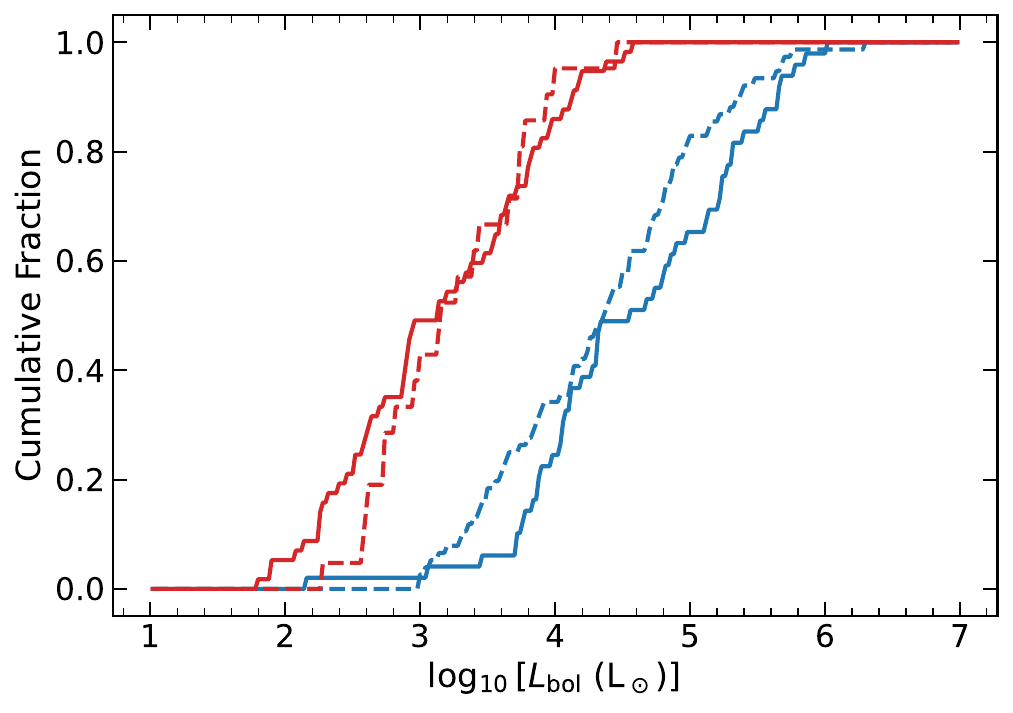}
    \includegraphics[width = 0.45\textwidth]{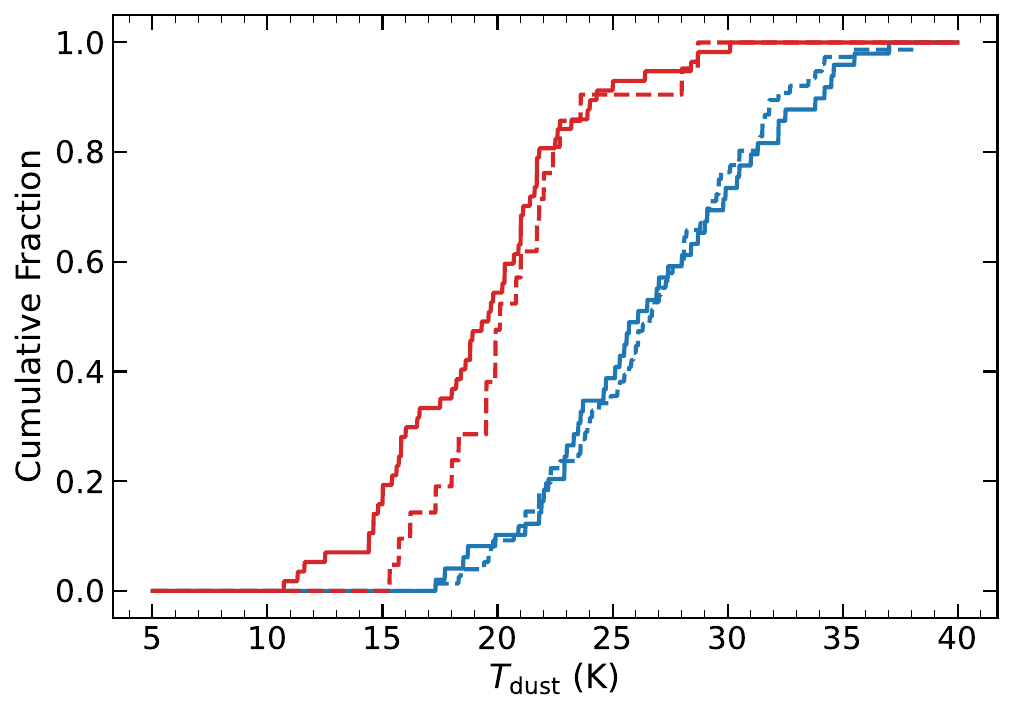}

    \caption{Cumulative distribution functions of bolometric luminosity (\textit{left}) and dust temperature (\textit{right}). Upper panels: comparison between our sample of Galactic clumps (solid curves) and the sample of \citet{Stephens2016} (dashed curves). Lower panels: comparisons between the objects classified as YSOs (red curves) and \hii regions (blue curves) drawn from our IRAM observations (solid lines)  and the \citet{Stephens2016} sample (dashed lines).}

    \label{fig:Stef_Us_cumulative}
\end{figure*}

In all but the upper left panel of Figure\,\ref{fig:V_disc_examples} we show examples of HCN, H$^{13}$CN, and CS spectra for seven sources with atypical HCN profiles. These plots illustrate the wide variation in emission profiles found throughout our sample. To avoid contaminating the sample and ensure accurate measurements of the clumps' HCN luminosity, profiles with strong evidence of self-absorption (e.g., see third row of panels) need to be identified and removed. For this purpose, we have chosen to compare the HCN spectral profiles with those of H$^{13}$CN and CS. H$^{13}$CN is an optically thin isotopologue of HCN and can be used to determine whether the HCN emission is affected by self-absorption by comparing the central velocities of the two tracers. However, because the H$^{13}$CN\,(1--0) transition is detected toward only \simm67\,per\,cent of our sample and, like HCN, presents with hyperfine structure, we use the CS\,(2--1) transition to identify cases where complexity in the HCN profiles may instead arise from multiple velocity components. CS is well suited for this purpose because it is a bright dense-gas tracer with a high detection rate (93\,per\,cent), a similar critical density to HCN\,(1--0) \citep[$4.7\times10^5\,{\rm cm^{-3}}$ and $1.3\times10^5\,{\rm cm^{-3}}$ for HCN\,(1--0) and CS\,(2--1), respectively;][]{Shirley2015}, and a simpler spectral profile without hyperfine structure.

We have grouped our sample based on their spectral profiles in the following way:

\begin{enumerate}[i]
    \item \textit{High confidence}: These are sources for which HCN has three resolved hyperfine components at the correct velocity separation between them and with relative intensities approximately in the 1:5:3 ratio (see top left panel of Figure\,\ref{fig:V_disc_examples}). Only 76 sources are classified as having highly reliable HCN spectra. \\

    \item \textit{Moderate confidence}: This category contains sources for which the spectral profile does not follow the expected relative intensity of the hyperfine lines or the lines appear blended. This effect is likely due to optical depth effects and/or non-local thermodynamic equilibrium (non-LTE) excitation conditions \citep{Loughnane2012}. The HCN, H$^{13}$CN, and CS measured velocities are in good agreement with each other and there is no evidence of self-absorption. We do not consider the complexity of these profiles to have a significant impact on the clumps' fluxes. There are 127 moderate confidence sources (see top right panel of Figure\,\ref{fig:V_disc_examples}). \\

    \item \textit{Multiple components}: Out of the full sample, 44 spectra contained multiple components, corresponding to separate clumps along the same line of sight (see the second row of Figure\,\ref{fig:V_disc_examples} for examples). For clumps with large velocity separation, we have separated the emission from the different components and additionally classified them into one of the previously mentioned categories. In these cases, we focus our analysis on the component that correlates with the velocity of the ATLASGAL source \citep{Urquhart2022}. We have successfully identified 22 additional clumps from this category, and the remaining 44 (22 that do not match ATLASGAL velocities and 22 whose components could not be resolved) are excluded from the sample. \\

    \item \textit{Self-absorbed}: These are sources for which the HCN line is distorted due to significant self-absorption and the flux measurement is no longer reliable. The majority of these present as four peaks due to an absorption feature in the central $F=2-1$ line. In more extreme cases, the HCN line becomes completely unrecognisable, displaying two distinct peaks separated by as much as $25\,\unit{\kilo\meter\per\second}$,  centred on the clump's systematic velocity (see the third row of Figure\,\ref{fig:V_disc_examples} for examples). The self-absorption is confirmed by the H$^{13}$CN line, whose central velocity aligns with the dip in HCN emission and generally agrees with the intrinsic velocity of the clump as determined by \citet{Urquhart2022}. In total, this affects 93 sources and these have been excluded from further analysis.\\

    \item \textit{Uncertain}: This last category is for any profiles where the HCN emission is considered unreliable but cannot be definitively classified as self-absorbed or having multiple components (see the fourth row of Figure\,\ref{fig:V_disc_examples} for examples). There are a total of 42 of these sources, and they are excluded from the analysis presented here.
\end{enumerate}

Following the classification process outlined above, we are left with 225 high and moderately reliable profiles (including the 22 clumps extracted from regions with multiple sources), where the fluxes are considered robust. While this filtering of clumps is not performed in extragalactic studies, it allows us to identify and separate populations that contribute differently to the observed line broadening and excitation conditions. Excluding sources with complex profiles therefore enables a cleaner comparison between physically distinct environments. Additionally, we show in Section\,\ref{HCN from diffuse gas} that the majority of HCN emission on cloud and larger scales originates from diffuse material, which is less sensitive to optical depth effects. Nevertheless, the remaining sample still constitutes a statistically significant set of sources spanning a broad range of physical conditions.

\subsection{Increasing the sample size}

\citet{Stephens2016} analysed the HCN emission towards 166 clumps observed as part of the MALT90 survey, providing an angular resolution (beam FWHM) of 38\,arcsec and a velocity resolution of 0.11\,\kms \citep{Jackson2013}. The MALT90 survey targets ATLASGAL sources, located in the 4th Quadrant ($300\degr <\ell < 350\degr$), and as such, their sample is highly complementary to ours. Combining the two samples not only increases the overall number of available clumps for analysis but also covers a larger range of Galactic longitudes. However, before doing this, we need to ensure that the samples are comparable.

The \citet{Stephens2016} sample is restricted to ATLASGAL sources with an associated IRAS counterpart, which limits their sensitivity to clumps hosting very luminous young high-mass stars. Their sample is, therefore, biased towards more evolved stages (i.e., massive YSOs and \hii regions) than our sample (see Section\,\ref{evolutionary classification} for details) and is therefore limited to sources with higher temperatures and luminosities (see top panels of Figure\,\ref{fig:Stef_Us_cumulative}). However, if we restrict our comparison to the YSO and \hii region samples (lower panels of Figure\,\ref{fig:Stef_Us_cumulative}), we find the agreement is much stronger, demonstrating the two samples are complementary and justifying our decision to combine them.

\begin{figure}
    \centering
    \includegraphics[width=0.48\textwidth]{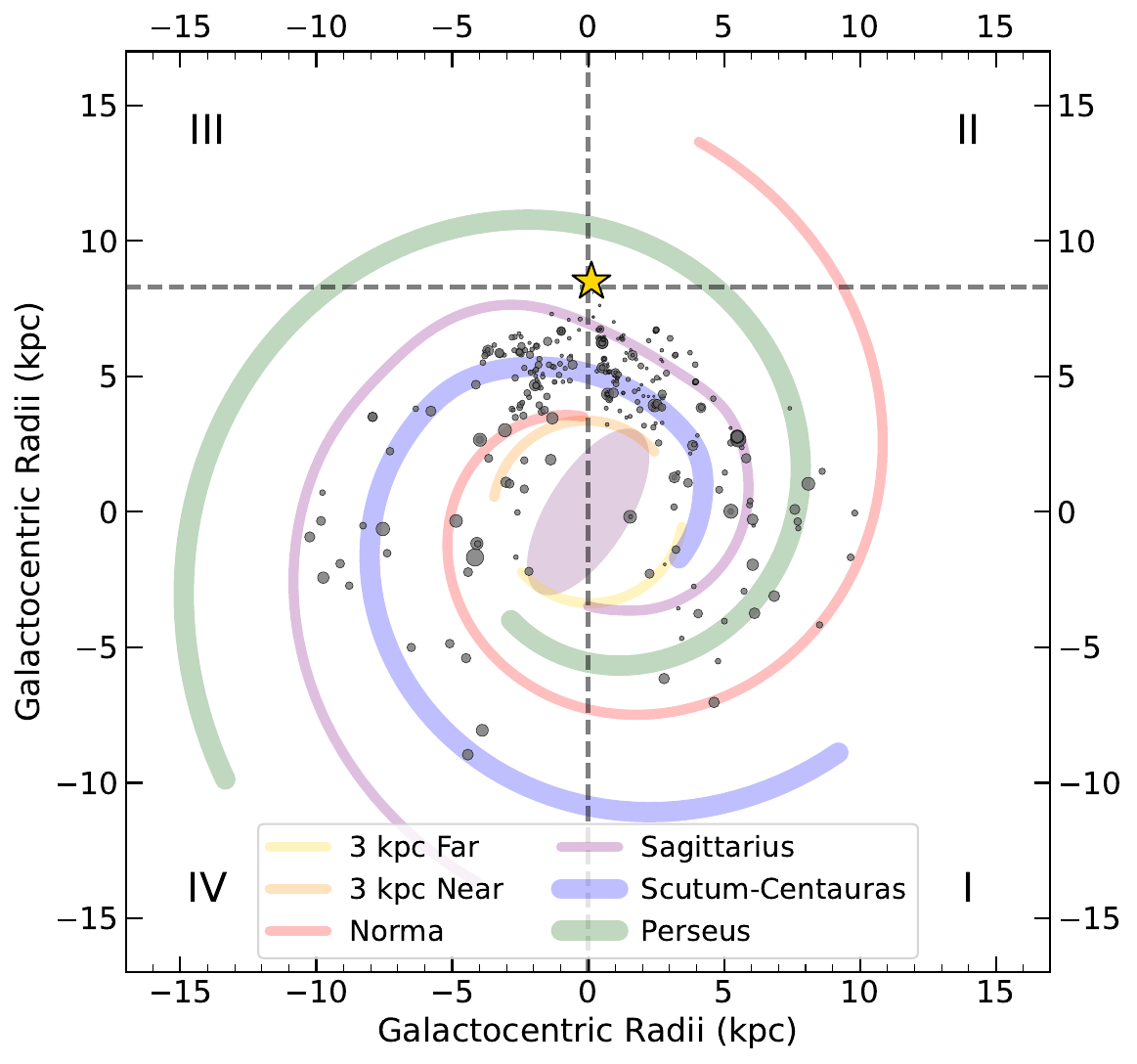}
    \caption{A schematic showing the loci of the Milky Way spiral arms according to the \citet{TaylorCordes1993} model, updated by \citet{Cordes2004} with the addition of a bisymmetric pair of arm segments representing the 3\,kpc arms. The distribution of our sample of 343 clumps is indicated by grey circles, the size of which is proportional to their bolometric luminosity. The star shows the position of the Sun, and the Roman numerals identify the Galactic quadrants. The light violet oval located in the centre of the diagram shows the position and orientation of the Galactic bar.}
    \label{fig:galactic distribution}
\end{figure}
To ensure consistency across the enlarged sample, we have visually inspected the HCN spectra from the \citet{Stephens2016} sample and applied the criteria described in the previous subsection. Forty-eight clumps are associated with strong self-absorption and have been removed from the sample. This leaves 118 high- or moderate- confidence sources, which are combined with our sample of 225 clumps to produce a science sample of 343 sources. The Galactic distribution of our sample can be seen in Figure\,\ref{fig:galactic distribution}.

\section{Characterising the Sample} \label{results}

\subsection{Evolutionary classification} \label{evolutionary classification}

\begin{figure}
    \centering
    \includegraphics[width = 0.45\textwidth]{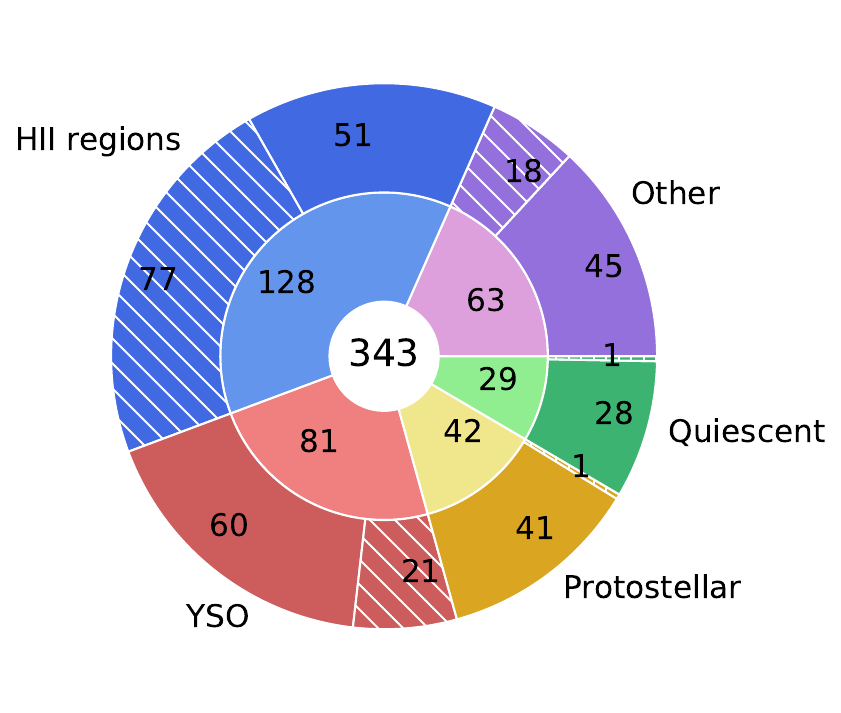}
    \caption{Distribution of the different clump classifications (see Section\,\ref{evolutionary classification}) within our combined sample of 343 clumps. The inner ring shows the distribution of classifications for the combined sample while the outer ring show how the individual classification breakdown between the IRAM sample (\textit{solid} coloured regions) and the \citet{Stephens2016} sample (\textit{hatched} regions).}
    \label{fig:pie}
\end{figure}

The ATLASGAL clumps have been classified through multi-wavelength analysis using mid- and far-infrared, submillimetre, and radio continuum maps (see \citealt{Konig2017} and \citealt{Csengeri2016} for details). Clumps are classified into four evolutionary stages as \hii regions, YSO, protostellar, quiescent, as well as clumps where a classification could not be reliably attributed (due to the ATLASGAL clump coinciding with significant amounts of extended mid-infrared emission), which in this paper we refer to as `Other' (\citealt{Urquhart2018,Urquhart2022}). Of the four evolutionary stages, only the first three actively form stars. The masses and densities of the quiescent clumps indicate they have the potential to evolve into star-forming clumps and some are indeed associated with molecular outflows indicating that star formation is already underway (\citealt{yang2018,yang2022}). This reflects the fact that clump evolution is a continuous process, and the division into discrete evolutionary stages is a necessary simplification for classification purposes rather than a reflection of sharp physical boundaries between them.

ATLASGAL clumps have sizes of \simm$0.5$-$1$\,pc and masses of \simm500\,M$_\odot$ and are therefore expected to be birthplaces of clusters rather than individual stars. As such, a single clump is expected to contain a number of protostellar objects with a range of different evolutionary stages. When that is the case, the clump's physical properties are assumed to be dominated by the most evolved embedded object, and the clump inherits the classification of that object (\citealt{Urquhart2018,Urquhart2022}).

Figure\,\ref{fig:pie} shows the distribution of clump classifications. Of the 343 clumps identified in the combined sample that are considered to have reliable fluxes, we find 280 are classified as one of the four star-forming evolutionary stages (we will call these our `evolutionary sample'). The remaining 63 clumps categorised as `Other' in Figure\,\ref{fig:pie} include clumps classified as photo-dominated regions \citep[PDR;][]{Jackson2013}, extended emission, ambiguous or complicated. The intensity and luminosity of the HCN emission has been measured for all 343 clumps (see Table\,\ref{tab:table}).

\begin{figure}
    \centering
    \includegraphics[width = 0.45\textwidth]{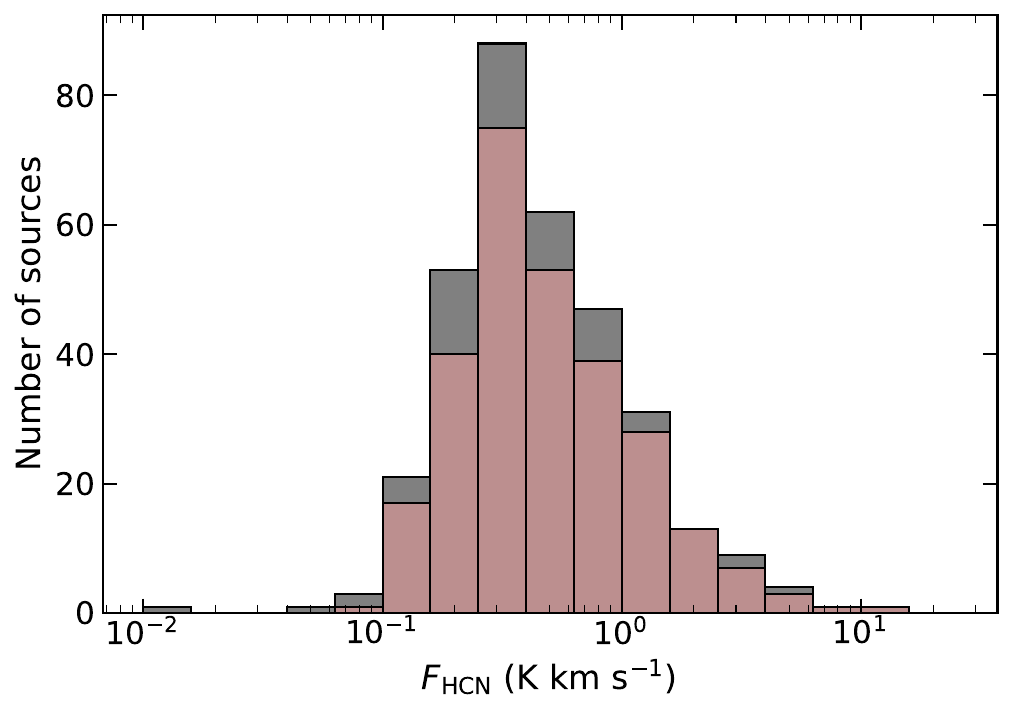}
    \caption{Distribution of $F_{\rm HCN}$ for all 343 clumps (\textit{grey}) in our science sample and the 280 clumps (\textit{rose}) classified in one of the four evolutionary stages in our evolutionary sample. The bin size is 0.2\,dex.}
    \label{fig:flux_hist}
\end{figure}

\subsection{Molecular Line Luminosity}

\begin{table*}
    \centering
    \caption{Main parameters for the full sample of 343 clumps. The distances, evolutionary classification and bolometric luminosities are adopted from the ATLASGAL catalogue. The HCN profile flags are described in Section\,\ref{data_red}.
}\label{tbl:clump_properties}
    \begin{tabular}{c . c c . . . . . c}
            \hline\hline
        ATLASGAL Name& \multicolumn{1}{c}{$\ell$} & \multicolumn{1}{c}{$b$} & Classification & \multicolumn{1}{c}{Distance} & \multicolumn{1}{c}{\vlsr}& \multicolumn{1}{c}{$L_{\text{HCN}}$} & \multicolumn{1}{c}{$\log_{10}[ L_{\text{bol}}]$} & \multicolumn{1}{c}{$\log_{10}[M_{\rm dust}]$}&\multicolumn{1}{c}{Profile Flag$^\star$}\\
        &\multicolumn{1}{c}{(deg)}&\multicolumn{1}{c}{(deg)}&  & \multicolumn{1}{c}{(kpc)} & \multicolumn{1}{c}{(km\,s$^{-1}$)}& \multicolumn{1}{c}{(K\,km\,s$^{-1}$)}  & \multicolumn{1}{c}{(\lsun)} & \multicolumn{1}{c}{(M$_\odot$)}& \\ \hline
        AGAL006.216$-$00.609&   6.216&$-$0.609&Protostellar      &3.0&18.5&12.3 &3.1& 2.6  &(i)\\
        AGAL008.049$-$00.242&	8.049&	$-$0.244&	Other&	5.0&	39.1	&6.8&	3.0&	2.4  &(i) \\
        AGAL008.706$-$00.414&   8.706&$-$0.413&Protostellar      &4.4&38.3&22.2&3.2&2.8 &(ii)\\
        AGAL010.104$-$00.416&	10.105&	$-$0.415&	Other&	2.9&	10.9&	1.8	&3.0	&	1.8    &(iii)\\
        AGAL010.214$-$00.306&	10.214&	$-$0.305&	Quiescent&	2.9&	11.3&	5.8&	2.8&	 2.1  &(ii)\\
        AGAL010.288$-$00.124&	10.286&	$-$0.121&	\hii region&	2.9&	12.6&	7.9&	3.7	& 2.4  &(ii)\\
        AGAL010.323$-$00.161&	10.322&	$-$0.160&	Other&	2.9&	12	&5.2&	5.3&	2.3  &(i)	\\
        AGAL010.342$-$00.142&	10.342&	$-$0.142	&Other&	2.9&	11.7&	2.4&	4.1&	2.4 &(ii)	     \\
        AGAL010.472$+$00.027&	10.472&	+0.028&	\hii region&	8.5&	67.1&	578.3&	5.6&	4.1  &(ii)\\
        \hline
    \end{tabular}
    \begin{tablenotes}
        \item Notes: Only a small portion of the data is provided here. The full table is available in electronic form at the CDS via anonymous ftp to cdsarc.u-strasbg.fr (130.79.125.5) or via http://cdsweb.u-strasbg.fr/cgi-bin/qcat?J/MNRAS/.
        \item $\star$ (i) high confidence, (ii) moderate confidence, (iii) multiple components, (iv) self-absorbed, (v) uncertain.
    \end{tablenotes}
    \label{tab:table}
\end{table*}

We have defined an emission line as having at least three consecutive channels above $3\sigma$ of the noise level.
The integrated line intensity is summed over all channels that satisfy the detection criterion. This has been used to calculate the molecular line luminosity, using the following equation (\citealt{Wu2005}):

\begin{equation}
    L_{\text{HCN}} = 23.5\times10^{-6}\,D^2\,\frac{\pi \left( \theta_{\text{s}}^2+\theta_{\text{beam}}^2\right)}{4\ln{2}} \int T_{\text{mb}}\,d\upsilon = D^2 F_{\rm HCN},
\end{equation}
\noindent where the kinematic distance $D$ to the source is in units of kpc, $\int T_{\text{mb}}\,d\upsilon$ is the velocity-integrated intensity in \unit{\kelvin\kilo\meter\per\second}, $\theta_{\text{s}}$ is the angular source size, and $\theta_{\text{beam}}$ is the beam size. Since we use single-pointing observations, we cannot estimate the source size, $\theta_{\text{s}}$, and instead we assume that the HCN emission filled the beam and therefore ${\theta_{\text{s}}^2 + \theta_{\text{beam}}^2} = 2\theta_{\text{beam}}^2$. For an easier conversion between flux and luminosity, $F_{\rm HCN}$ is essentially $L_{\rm HCN}D^{-2}$ (this is the same as the convention used by \citealt{Stephens2016}).

We have used this method to determine the molecular line luminosity for all sources that fulfil the 3$\sigma$ threshold and have kinematic distances, which were taken from  \citet{Urquhart2022}. In Figure\,\ref{fig:flux_hist}, we show the distribution of $F_{\text{HCN}}$ for our previously defined science sample of 343 clumps and the evolutionary sample of 280 clumps. Five sources classified as one of the four evolutionary stages do not have a reliable distance available and will therefore not be included in the analysis that follows.

The observations used by \citet{Stephens2016} have a beam size of 38\,arcsec. To measure the total flux of their clumps, they have fitted their emission maps with a 2D Gaussian and integrated the emission over the full area of the fit. However, our sample is made up of single-beam observations and therefore our method for measuring the flux for our clumps differs. To maintain consistency between the samples we have chosen to use the peak flux from the moment-0 maps for the objects in \citet{Stephens2016} and thus make all sources in our combined sample beam-sized. The difference in flux between the two samples is consistent with what we would expect from the different beams. This allows us to make a more direct comparison to clump properties derived from the ATLASGAL 870-\micr dust emission maps.

\subsection{Correlation between bolometric and infrared luminosities}
\label{IRvBOL}

\begin{figure}
    \centering
    \includegraphics[width = 0.45\textwidth]{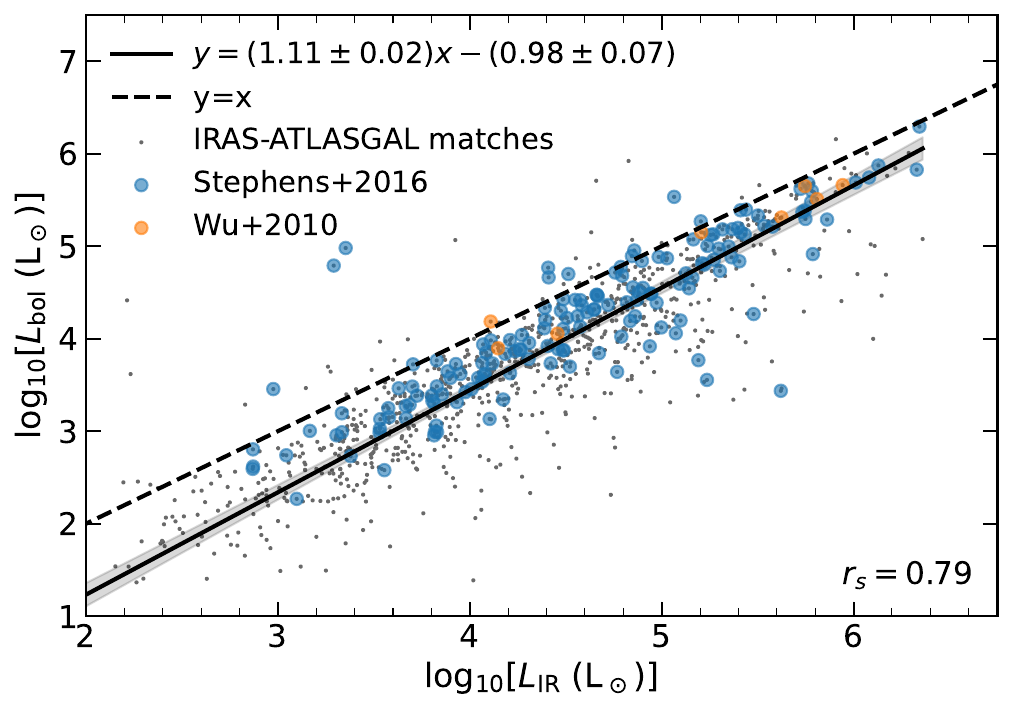}
    \caption{Comparison of the ATLASGAL  bolometric and IRAS IR luminosities. Targets observed by \citet{Wu2005} and \citet{Stephens2016} are highlighted in orange and blue, respectively. The dashed line is a line of equality. The solid line is a fit to the complete sample of ATLASGAL sources with an IRAS counterpart, and the uncertainty is represented by a shaded region around the fit. The correlation coefficient is shown in the bottom right.}
    \label{fig:Bol vs IR}
\end{figure}

The ATLASGAL catalogue provides bolometric luminosities; however, most previous studies use infrared luminosities calculated from IRAS fluxes (e.g., \citealt{Gao+Solomon2004}, \citealt{Wu2005}, \citealt{Stephens2016}). To compare our sample with previous studies, we derive a conversion factor between the bolometric luminosities ($L_{\text{bol}}$) and infrared luminosities ($L_{\text{IR}}$).
Previous works relating infrared and HCN luminosities estimated the total infrared luminosity (8–1000\,\micr) using an empirical extrapolation from the four IRAS photometric bands (12, 25, 60, 100\,\micr; \citealt{Sanders&Mirabel1996}):

\begin{equation}
    L_{\rm IR} = 0.56D^2(13.48f_{12}+ 5.16f_{25} +2.58f_{60}+ f_{100})
\end{equation}
where the flux densities, $f_n$, are in units of Jy, $L_{\rm IR}$ is in units of solar luminosity (L$_\odot$) and the distance (D) is in kpc. This relation is calibrated on largely unresolved external galaxies and may not accurately capture the spectral energy distribution (SED) shape of individual, more deeply embedded Galactic clumps. In contrast, ATLASGAL bolometric fluxes are derived by directly integrating an SED fit to multi-wavelength photometry spanning 8–870\,\micr\ \citep[for more information on the fitting method, refer to][]{Konig2017}.

To derive the conversion factor, we cross-matched the ATLASGAL CSC with the IRAS Point Source Catalogue using a 30\,arcsec search radius, resulting in \simm1200 matched sources with both bolometric and infrared luminosity estimates. Since luminosities depend on the adopted distances, and previous studies typically relied on kinematic distances derived independently for each source, using different Galactic rotation curve models, the infrared luminosities were recalculated using the ATLASGAL distances \citep{Urquhart2022} to ensure both bolometric and infrared measurements are placed on a single, consistent distance scale. In Figure\,\ref{fig:Bol vs IR}, we compare the bolometric and infrared luminosities for the matched sources, and highlight those that have been targeted by \citet{Wu2005} and \citet{Stephens2016}. The plot reveals a strong correlation between the two quantities (Spearman correlation coefficient of 0.79 and $p$-value $\ll 0.0013$). An orthogonal regression fit returns a slope of $1.11\pm0.02$, indicating an approximately linear relationship between the two luminosities.

Separating the matched sources by evolutionary stage yields consistent relationships; however, the numbers of matched quiescent and protostellar sources are too small for statistically meaningful independent fits. The IRAS infrared luminosities are typically a factor of \simm2--3 larger than the bolometric luminosities for the matched sample. This is despite the infrared luminosity being derived from a narrower wavelength range (12–100\,\micr) than the bolometric luminosity (8–870\,\micr). We attribute this offset primarily to the IRAS beam size ($\sim$0.5--3\,arcmin; \citealt{IRAS_Resolution_2005}), which is significantly larger than the typical size of ATLASGAL clumps ($\sim$1\,arcmin; \citealt{Contreras2013}), likely resulting in blended flux from nearby sources.

We have applied this conversion factor to our ATLASGAL sample of HCN detections to estimate their expected IRAS infrared luminosities. This allows us to investigate the relationship between IR luminosity and HCN luminosity, while at the same time taking advantage of the large sample of clumps covered by the ATLASGAL catalogue and extend the previous analysis to a wider range of evolutionary stages. While the above conversion between bolometric and IRAS infrared luminosity is primarily derived for \hii\ regions and YSOs, we apply it to our full sample (including earlier evolutionary stages) to allow a direct comparison with the previously established IR-HCN relationship. We note that quiescent clumps, being IR-dark, do not strictly possess an IR luminosity due to the absence of an embedded 70\,\micr point source, and the derived value should merely be treated as a nominal quantity, scaled directly from the bolometric luminosity.

\subsection{HCN as a tracer of mass} \label{HCN as a mass tracer}
\begin{figure}
    \centering
    \includegraphics[width = 0.45\textwidth]{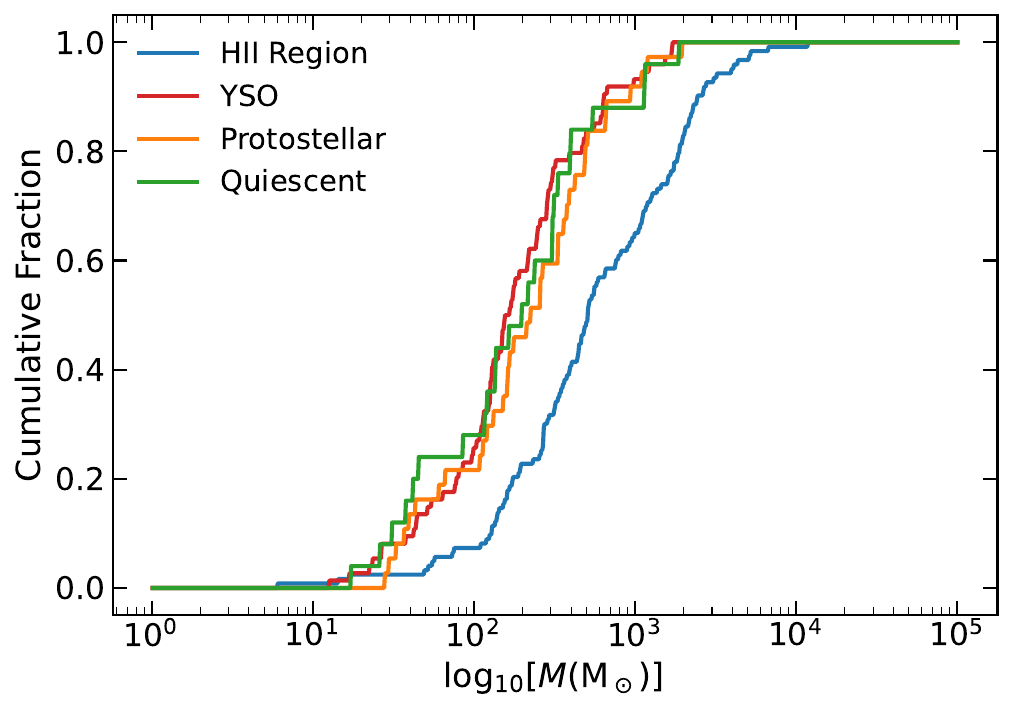}
    \includegraphics[width = 0.45\textwidth]{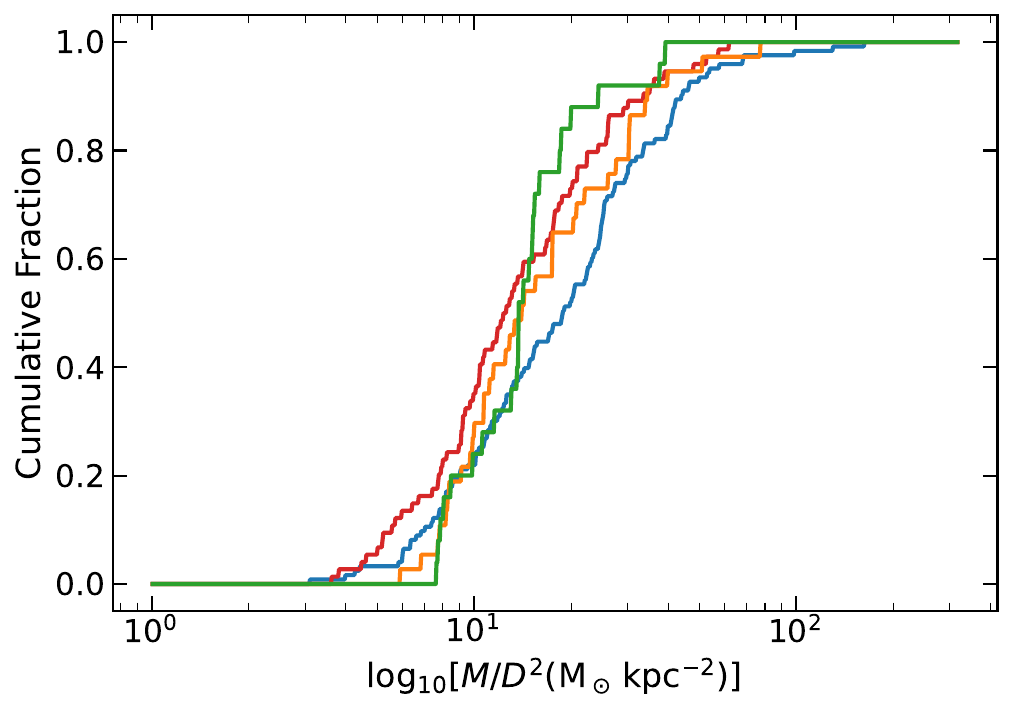}
    \caption{Cumulative distribution functions of the mass within the beam (\textit{top panel}) and the peak surface density (\textit{bottom panel}) for the different evolutionary stages (see legend for details).}
    \label{fig:CDF_Mass}
\end{figure}

HCN is a commonly used dense gas tracer and one of the most prominent molecular lines in extra-galactic studies \citep[e.g.,][]{GraciaCarpio2008, Usero2015, Aladro2015,Bigiel2016, Neumann2025}. We have calculated the 870-\micr flux within a 30- and 38-arcsec aperture directly from the ATLASGAL emission maps (\citealt{ATLASGAL2009, Csengeri2014}) and used these values to derive the total mass within the beam for our and the \citet{Stephens2016} samples, respectively.  This ensures that we are comparing the dust mass and HCN luminosities within the same aperture. Clump masses were derived from the 870-\micr\ dust continuum flux assuming optically thin emission, following \citet{Hildebrand1983}. The adopted dust opacity of $\kappa_{870}=1.85\,{\rm cm^2g^{-1}}$ is calculated as the average of the \citet{OssenkopfHenning1994} dust models for a dust emissivity index of $\beta = 1.75$, and a gas-to-dust mass ratio of 100. Dust temperatures are derived for each source from modified-blackbody SED fitting to the ATLASGAL and Hi-GAL photometry (\citealt{Konig2017}). The typical mass uncertainty arising from these assumptions is of order 20\,per\,cent, dominated by the uncertainties on dust temperature and integrated flux \citep{Urquhart2018}. In the upper panel of Figure\,\ref{fig:CDF_Mass}, we show the cumulative distribution function of the mass within the beam for each evolutionary stage. This reveals \hii regions have significantly higher masses compared to the earlier stages. The lower panel of Figure\,\ref{fig:CDF_Mass} shows the peak surface density, which shows all evolutionary stages have similar distributions. A similar pattern is seen for $L_{\rm HCN}$ and $F_{\rm HCN}$. This means that the higher masses found for \hii regions are the result of a distance bias, i.e., \hii regions are very luminous and are detected at larger distances than the other evolutionary stages.

\begin{figure}
    \centering
    \includegraphics[width = 0.45\textwidth]{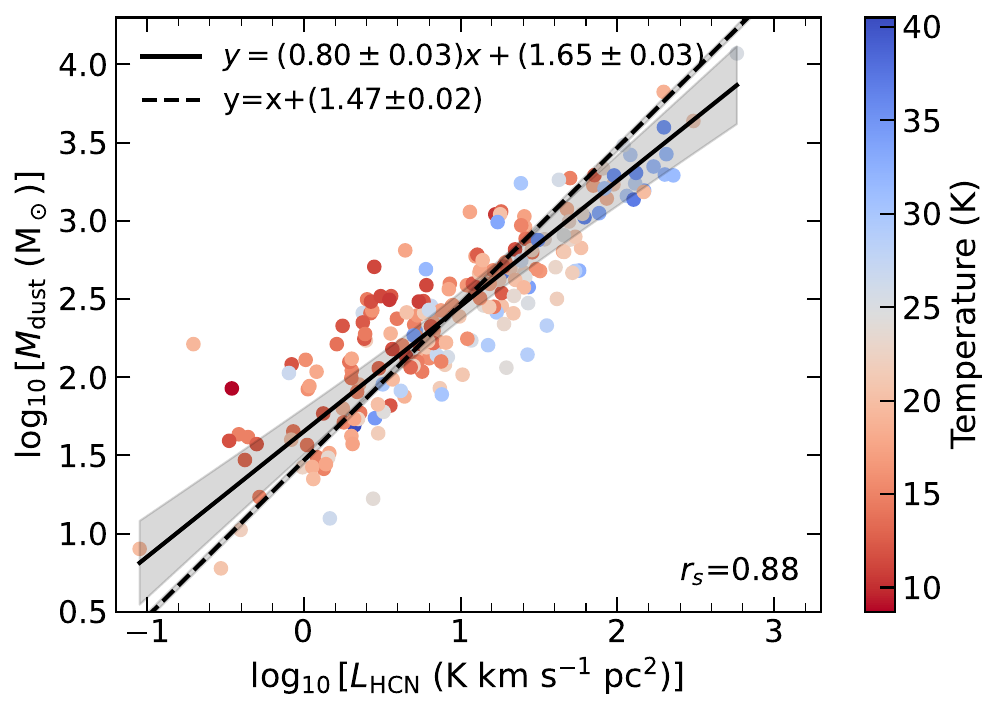}
    \caption{Distribution of the $L_{\rm HCN}$ and $M_{\rm dust}$ for our sample of 343 clumps. Each point has a different colour depending on the dust temperature of the clump, with red and blue colours indicating cooler and hotter temperatures respectively. The orthogonal fit is shown as a \textit{solid black} line with the shaded area around it representing the $1\sigma$ uncertainty in the slope, while the fit assuming a linear conversion (i.e. slope of one) is shown as a \textit{dashed black} line. The partial Spearman correlation coefficient, with temperature as the controlled variable, is shown in the bottom right corner.}
    \label{fig:HCN_M}
\end{figure}

In Figure\,\ref{fig:HCN_M}, we show the correlation between dust mass, HCN luminosity and the dust temperatures (\citealt{Urquhart2022}) for our sample of 343 clumps. The fit to the full sample is $0.80\pm0.03$ and the partial-Spearman correlation coefficient \citep{partial_spearmann} of $r_s=0.88$ and $p$-value $\ll0.0013$ (using temperature as the controlled variable) suggests this correlation is independent of temperature.

Many previous studies have assumed a linear relationship between mass and HCN luminosities and adopted a conversion factor of 10, i.e., $M\,[\rm M_\odot] = 10\,L_\mathrm{HCN}\,[{\rm K\,km\,s^{-1}\,pc^2}]$ (e.g.,\,\citealt{Gao+Solomon2004, Usero2015,Gallagher2018,Jimenez-Donaire2019}; also see \citealt{TafallaUseroHacar2023} who provide a detailed discussion of the variance in this value). Given that our fit is close to unity and for easy comparison with the literature we have refitted the data with a unit-slope regression, resulting in the following relationship $M = (29.5\pm1.4)\times L_{\rm HCN}\,{\rm M_\odot}$, which is significantly higher than the general value used but is consistent with the upper limit suggested by \citet{Gao+Solomon2004}. The mass conversion factor has been shown to vary by almost an order of magnitude for metallicities between 0.4\,Z$_\odot$ and 1.3\,Z$_\odot$ \citep{Patra2025} with a conversion factor of \simm20 at solar metallicity. Nevertheless, this allows us to estimate the $L_{\rm HCN}$ of clumps with known masses and we will use these values in Section\,\ref{HCN from diffuse gas}.

\section{Exploring the star formation relation}
\label{SF relation}

\begin{figure}
    \centering
    \includegraphics[width = 0.45\textwidth]{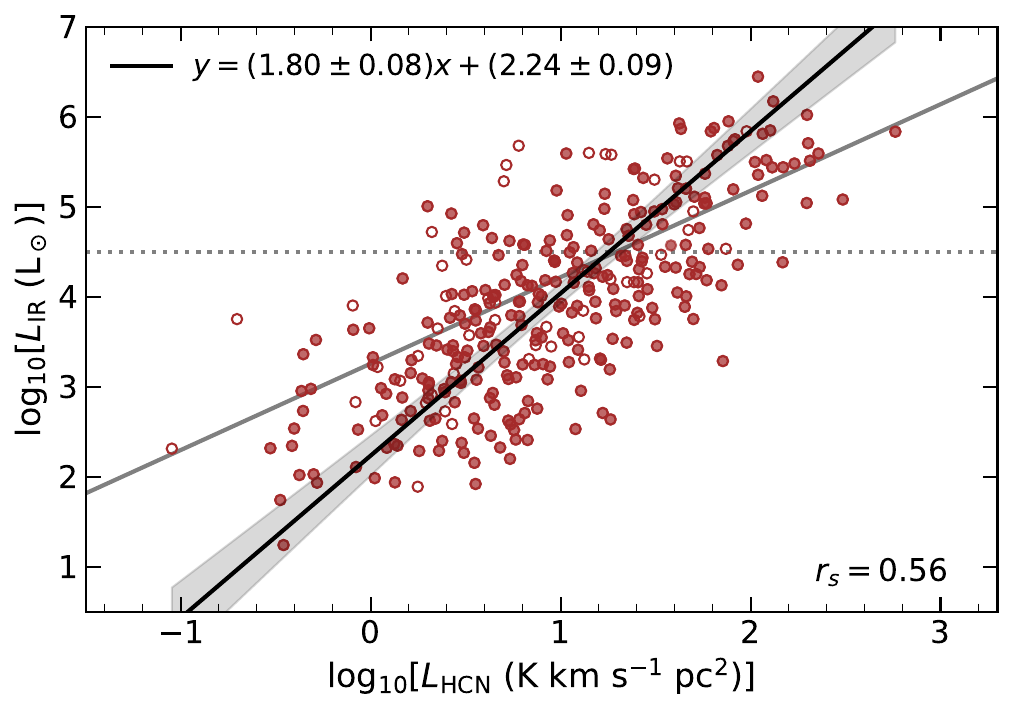}
    \caption{Total IR luminosity against HCN line luminosity for our sample of 343 sources, with clumps that are not classified as part of our evolutionary sample marked with open circles. The grey line is the fit to the \citet{Wu2010} and \citet{Gao+Solomon2004} samples discussed in Section\,\ref{Introduction} (i.e.,  of $0.96\pm0.01$) and the dotted line marks $L_{\rm IR}= 10^{4.5}\,{\rm L_\odot}$. The orthogonal regression fit for the sample is shown in black, with the shaded region representing the 1$\sigma$ uncertainty to the fit. The partial Spearman correlation coefficient, with distance as the controlled variable, is shown in the bottom right corner.}
    \label{fig:combined_IR_HCN}
\end{figure}

A goal of this work is to extend the star formation relation demonstrated by \citet{Wu2005} to a larger and more representative sample of Galactic clumps and investigate how the relationship might change for different evolutionary stages. In Figure\,\ref{fig:combined_IR_HCN}, we show the IR luminosity and HCN line luminosity for our sample of 343 clumps. The distribution of the whole sample reveals a moderate correlation between the two parameters (with a correlation coefficient of 0.56 and p-val$\ll0.0013$) and an orthogonal fit with a slope of $1.80\pm0.08$. However, we note that this deviates significantly from the \citet{Wu2005} and \citet{Gao+Solomon2004} relation.

This raises the question: do all clumps indeed follow the same star-formation law as seen in nearby galaxies? \citet{Wu2005} reported a deviation from a linear correlation for clumps below $L_{\rm IR} > 10^{4.5}$\,\lsun, when the slope becomes steeper; however, we do not observe the same in our data. The apparent `turn-off' for the \citet{Wu2005} sample was a result of the smaller sample size and the bias towards very luminous star-forming clumps. In fact, consideration of the full sample of clumps yields a steeper slope of $1.46\pm0.07$. Additionally, applying the same luminosity cut to our sample, namely $L_{\rm IR} > 10^{4.5}$\,\lsun\ (dotted line in Figure\,\ref{fig:combined_IR_HCN}), yields a slope consistent with their reported value.

In the analysis presented in the rest of the paper, we focus on the 275 ATLASGAL clumps for which we have reliable distances and have been classified into one of the four evolutionary stages associated with star formation activity, with quiescent clumps representing the precursor phase.

\subsection{Correlation between HCN and IR luminosities}
\label{Gao-Solomon Relation}

\begin{figure*}
    \centering
    \includegraphics[width = 0.9\linewidth]{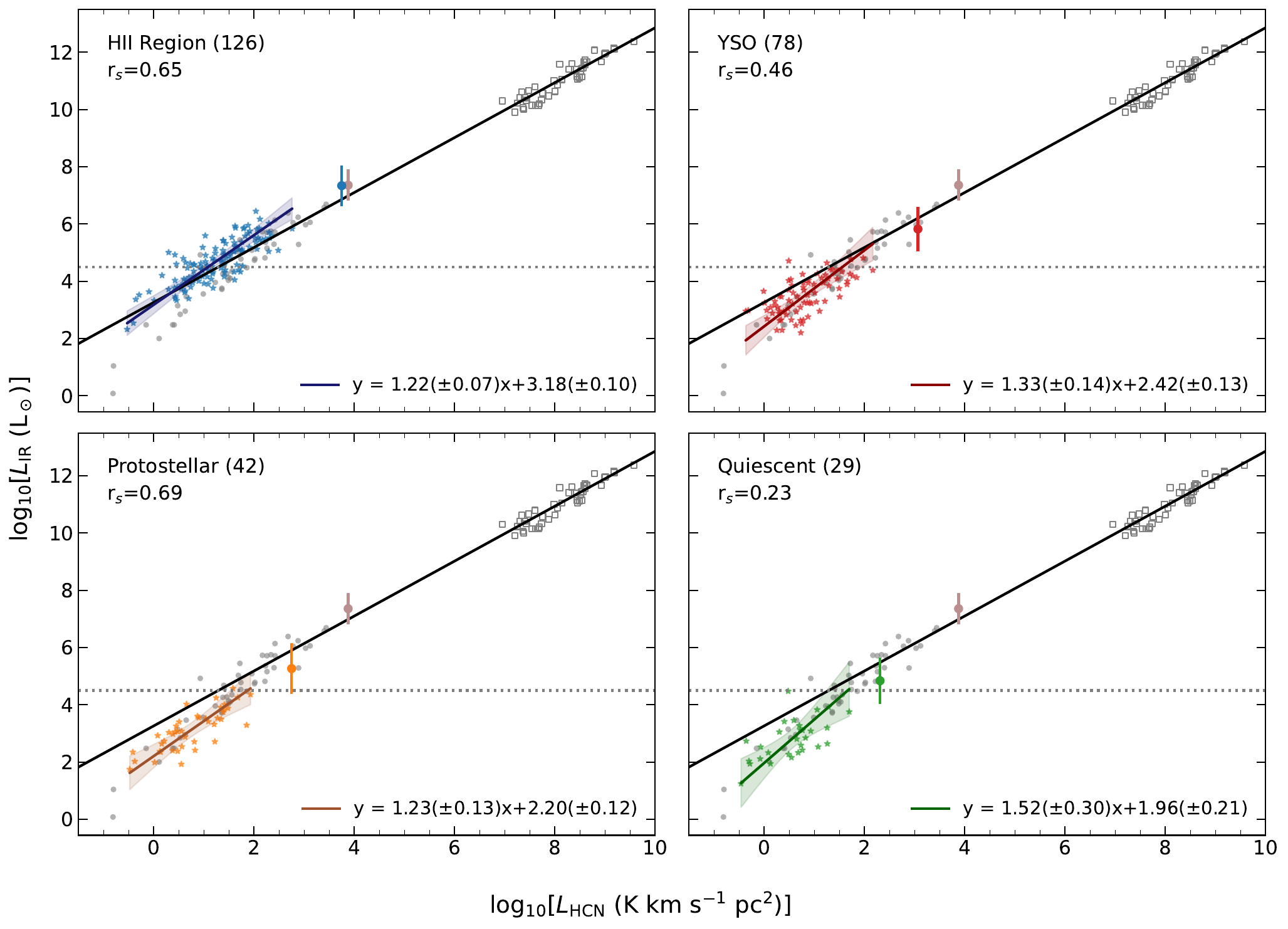}
    \caption{Correlation between $L_{\rm HCN}$ and $L_{\rm IR}$ for the 4 evolutionary stages identified. The grey markers (same in each panel) show the extra-galactic data from \citet{Gao+Solomon2004} (open squares) and the clump data from \citet{Wu2010} (closed circles). Coloured stars show data from the ATLASGAL survey and each colour corresponds to a different evolutionary stage (identified at the top left of each panel together with the sample size in brackets). The black line through the grey points is limited to objects with $L_{\rm IR}>10^{4.5}$\,L$_\odot$, following the method in \citet{Wu2005} and has a slope of $y=0.96x+3.26$. The ATLASGAL data have also been fitted, with each colour corresponding to the respective evolutionary stage. The shaded area around each coloured fit represents the $1\sigma$ error of the fit. The partial Spearman correlation coefficient of ATLASGAL data, with distance as a controlled variable, is shown at the top left of each panel. The total luminosity of each stage is shown as a filled circle and error bar with the corresponding colour and the total luminosity of the full sample of 275 clumps is marked as a rosy-brown point. The error bars on those points show the uncertainty, which is dominated by the uncertainty in the conversion between bolometric and infrared luminosity (the uncertainty in $L_{\rm HCN}$ is well contained by the size of data points).}
    \label{fig:L_hcn v L_ir}
\end{figure*}

\begin{figure}
    \centering
    \includegraphics[width = 0.45\textwidth]{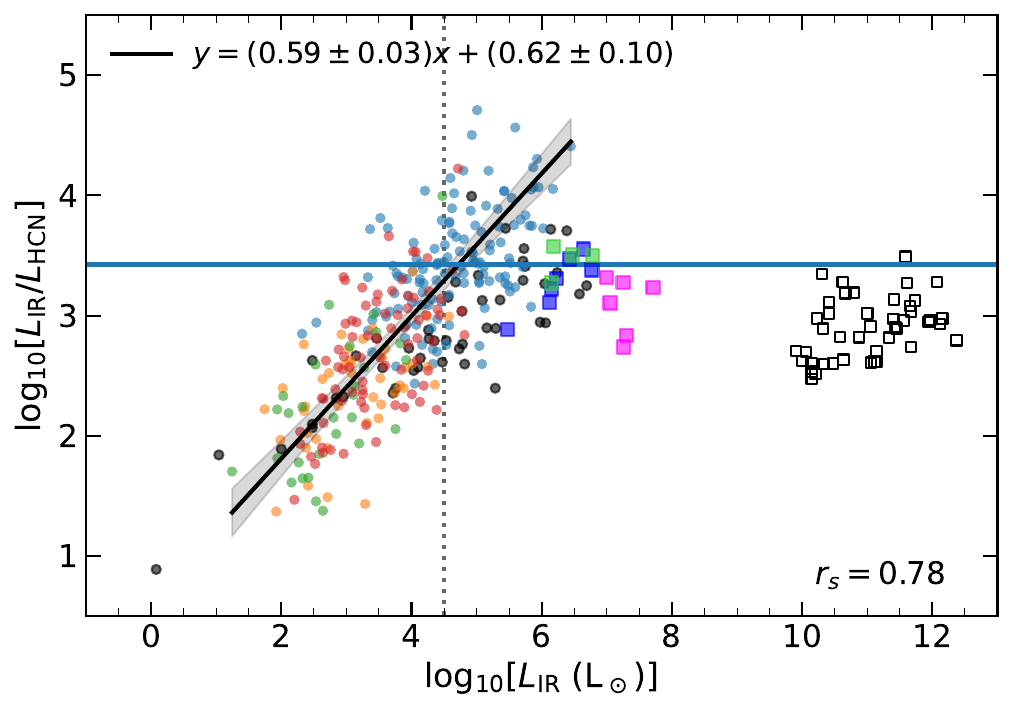}
    \caption{Distance-independent ratio of $L_{\text{IR}}/L_{\text{HCN}}$ against $L_{\text{IR}}$ for our sample of 275 clumps (colour-coded according to their evolutionary stage as in Figure\,\ref{fig:L_hcn v L_ir}). The sample of Galactic clumps by \citealt{Wu2010} (solid black circles) and the sample of nearby galaxies of \citealt{Gao+Solomon2004} (open black squares) are also shown. The fit to the Galactic clumps is shown in black with the $1\sigma$ uncertainty represented by the shaded region around the line. The horizontal blue line marks the mean value of $L_{\text{IR}}/L_{\text{HCN}}$ for our sample, while a dotted grey line marks $L_{\rm IR}=10^{4.5}\,{\rm L}_\odot$. Squares show extra-galactic GMCs for comparison, same as in Figure\,\ref{fig:Wu_slopes} \citep{Rosolowsky2011, Buchbender2013, Chen2017}. The partial Spearman correlation coefficient, using distance as the control, for our sample is shown in the bottom right corner.}
    \label{fig:ratio_ir}
\end{figure}

To further investigate the relationship for different stages of star formation, we have replicated the results of \citealt{Gao+Solomon2004} and \citet{Wu2010} in Figure\,\ref{fig:L_hcn v L_ir}  and included our evolutionary sample of 275 ATLASGAL clumps. Each panel in Figure\,\ref{fig:L_hcn v L_ir}, tracks one of these four stages (quiescent, protostellar, YSO, and \hii region) associated with star formation \citep{Urquhart2022}. This figure shows how each evolutionary stage of our sample compares to the previous studies. A partial Spearman correlation coefficient has been calculated for each evolutionary stage, with distance as the controlled variable. The \hii regions, YSOs, and protostellar samples have strong or moderate correlations (i.e., $r_s$ of 0.65, 0.46, 0.69, respectively, with  $p$-values $\ll0.0013$) while the quiescent sample is uncorrelated ($r_s=$ 0.23 with a $p$-value $=0.23$), which is consistent with the absence of a strong infrared source in these clumps. The slopes of the three actively star-forming stages are broadly consistent with each other \citep[and the slope found by ][]{Stephens2016}, yet significantly different from the slope of $0.96\pm0.01$ demonstrated in Figure\,\ref{fig:Wu_slopes} using a restricted sample. Additionally, the steeper slope of the whole sample seems to be an effect of the different interception points (rather than the slopes) of the subsamples, caused by a disproportionate increase in IR luminosity compared to the line luminosity.

We find that approximately two-thirds of \hii regions lie above $L_{\rm IR}= 10^{4.5}$\,\lsun, which was set as a threshold by \citet{Wu2005} in their fit. They are most closely aligned with the previously established slope of $0.96\pm0.01$ (see Figure\,\ref{fig:L_hcn v L_ir}), which is not surprising, as the previous studies have mainly focused on more evolved clumps. However, almost all star-forming clumps at earlier stages are located below the established relationship and the $10^{4.5}$\,\lsun\ threshold.

In Figure\,\ref{fig:L_hcn v L_ir}, we also show the total $L_{\text{IR}}$ and $L_{\text{HCN}}$ for all clumps in each stage (these are shown as large filled circles) and the totals for the entire sample (large filled rose coloured circles). Comparing the total values of the four evolutionary samples to the total values of the whole sample, we find the \hii-region sample provides most of the HCN and IR luminosity. Despite \hii regions making up only 45\,per\,cent of the sample, they contribute  74\,per\,cent of the total HCN luminosity and 96\,per\,cent of the total $L_{\text{IR}}$. In fact, with a mean IR luminosity of $\log_{10}( L_{\text{IR}})=5.24\,{\rm L}_\odot$, \hii regions are on average 20 times brighter than the preceding stage (YSOs) and with a mean $\log_{10}(L_{\text{HCN}})=1.65\,{\rm K\,km\,s^{-1}\,pc^2}$, are three times brighter than YSOs and protostellar clumps on average and six times brighter than quiescent clumps (see Table\,\ref{tab:means}).

\setlength{\tabcolsep}{4pt}
\begin{table}
    \centering
    \caption{A table containing the mean and total IR and HCN luminosities for each of our sub-samples.}
    \begin{tabular}{lcccc}
     \hline \hline
          & $\log_{10}(\bar{L_{\rm IR}})$ & $\log_{10}(\bar{L_{\rm HCN}})$&  $\log_{10}(\Sigma L_{\rm IR})$& $\log_{10}(\Sigma L_{\rm HCN})$\\ \hline
        \hii regions   &5.24\,(0.07) & 1.65\,(0.06)& 7.34\,(0.71)& 3.75\,(0.05)\\
         YSOs         & 3.93\,(0.07)& 1.16\,(0.06)& 5.83\,(0.77)& 3.05\,(0.05)\\
         Protostellar & 3.64\,(0.11)&1.13\,(0.09)& 5.26\,(0.89)& 2.75\,(0.05)\\
         Quiescent    & 3.38\,(0.13)& 0.85\,(0.10)& 4.84\,(0.82)& 2.31\,(0.05)\\ \hline
         All          &4.92\,(0.06) &1.44\,(0.04) &7.36\,(0.55) &3.88\,(0.03)\\
         \hline
    \end{tabular}
    \begin{tablenotes}
        \item{Note:  Uncertainties are shown in brackets. The uncertainty in the mean is the standard error of each sample.}
    \end{tablenotes}

    \label{tab:means}
\end{table}

\setlength{\tabcolsep}{6pt}

The fact that the infrared luminosity is dominated by the \mbox{\hii\ regions} is not surprising given the bolometric luminosity of embedded high-mass stars changes over 3-4 orders of magnitude during their evolution. However, what is more surprising is the contribution \hii\ regions make to the total HCN luminosity given that previous studies show that the clump mass does not change significantly with each stage \citep{Urquhart2022}.

\subsection{Is the infrared-HCN luminosity relation universal on all scales?}

Since stars form from dense gas, it is natural to assume that if we have a measurement of the dense gas mass within a clump or galaxy, we should be able to determine the SFR with a simple assumption of star-formation efficiency. With the prospect of a universal star formation law, \citet{Wu2005} turned to HCN as a good dense-gas tracer, for which there is already an established relationship with SFR for nearby galaxies \citep{Gao+Solomon2004}. Indeed their sample of bright \hii regions seems to align well with the extragalactic relationship. However, as we have shown in the previous section, that does not seem to be the case when we consider a more complete sample that includes less evolved clumps.

To investigate this in more detail, in Figure\,\ref{fig:ratio_ir}, we plot the $L_{\rm IR}/L_{\rm HCN}$ ratio, which is a distance-independent parameter, with respect to the IR luminosity. If the relationship between dense gas traced by HCN and SFR is a universal law on all scales and for all star-forming stages, we would expect the ratio of the luminosities to be constant with respect to $L_{\text{IR}}$. However, we find that this ratio increases as clumps evolve (a similar behaviour is also observed by \citet{Ma2013} for HCO$^+$). In this plot, we also show the mean value of $L_{\text{IR}}/L_{\text{HCN}}$ for all of our clumps as a blue horizontal line, which is on the higher end of extra-galactic observations, due to the total luminosity of our sample being dominated by \hii regions. It is clear that the relationship, previously described as universal, between IR and HCN luminosity, only holds true for the brightest regions and not for the full range of clumps. This disconnect poses a problem for the interpretation of the well established extra-galactic relationship as \hii regions only represent the final stage of star formation, once the central star has already been formed.

\section{Implications for the Milky Way} \label{sect:milkyway}

The Milky Way is a spiral galaxy and is morphologically similar to the sample of normal spiral galaxies investigated by \citet{Gao+Solomon2004}. As such, it is reasonable to assume that, if viewed by an external observer, it would have total IR and HCN luminosities similar to those of other nearby late-type galaxies. If this is the case, we should be able to estimate how many clumps are needed for our Galaxy to produce these values.

\citet{Stephens2016} used their results to estimate the total IR and HCN luminosity of the Milky Way and argue that there is not enough HCN emission from dense clumps to satisfy the Gao-Solomon relation as hypothesised by \citet{Wu2005}. They show that, if the HCN emission originated only from dense gas in high-mass star-forming clumps, our Galaxy would need \mbox{$\sim 1$-$2 \times 10^5$} high-mass clumps ($M_{\rm clump}>1000$\,M$_\odot$) to be comparable to the sample of spiral galaxies observed by \citet{Gao+Solomon2004}. This was approximately an order of magnitude more than the number known at the time (e.g., \citealt{Urquhart2014}), leading them to suggest that the majority of HCN luminosity must come from more diffuse gas within galaxies. Since then, the \higal survey (Herschel InfraRed Galactic Plane Survey; \citealt{Hi-GAL2010}) has catalogued $\sim150\,000$ dense clumps, covering the whole of the Galactic mid-plane (\citealt{Elia2021}; 360\degr\ in $\ell$ and 2\degr\ in $b$). It is, therefore, timely to revisit the analysis performed by \citealt{Stephens2016} and reassess the contribution of sub-thermal emission to the total Galactic HCN luminosity.

\subsection{Estimating the HCN and IR luminosity of the Milky Way} \label{estimating total luminosities}

The \higal survey \citep{Molinari2008} surveyed the whole Galactic mid-plane and has produced the most complete census of dense clumps in the Milky Way \citep[\simm 150\,000 clumps;][]{Elia2021}. Although Hi-GAL provides approximately the right quantity of clumps hypothesised by \citet{Stephens2016}, the majority of them are quiescent (i.e., lacking a 70\,\micr detection) and so it is unlikely that the HCN emission from dense clumps alone is sufficient to make the Milky Way comparable to galaxies in the \citet{Gao+Solomon2004} sample. We do not have HCN line observations of the majority of the \higal clumps, however, we can use the strong correlation found in Section\,\ref{HCN as a mass tracer} to convert the clump masses to HCN luminosities and integrate the emission to obtain an estimate for the total HCN luminosity contributed by dense gas.

\begin{figure}
    \centering
    \includegraphics[width = 0.45\textwidth]{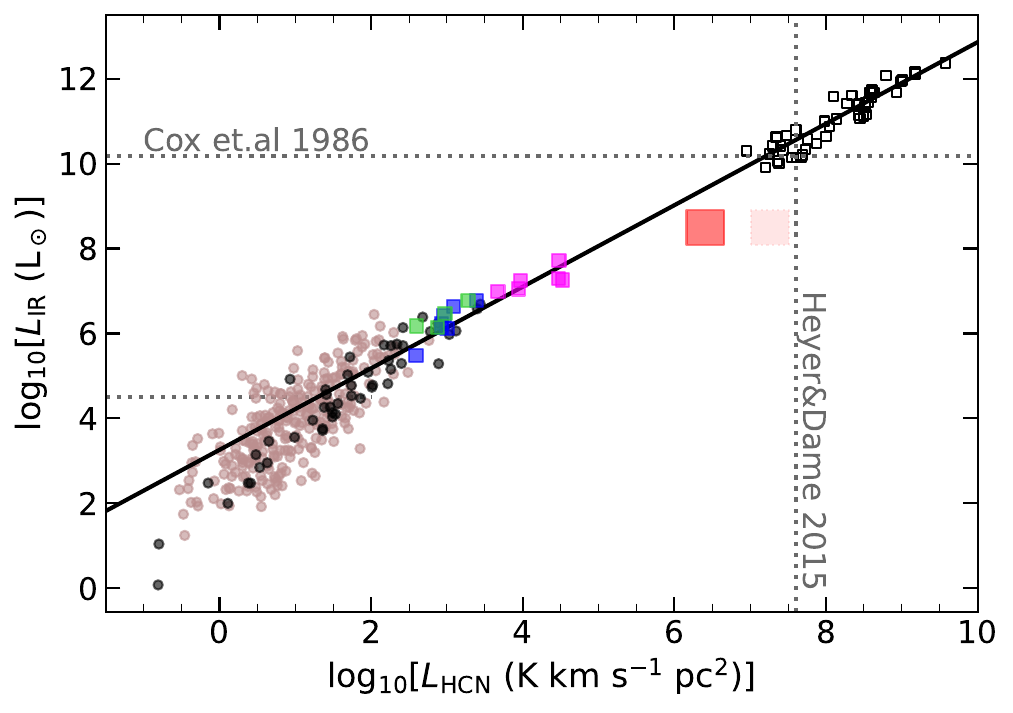}
    \caption{Distribution of $L_{\rm HCN}$ versus $L_{\rm IR}$ for the full sample of 275 clumps presented in this work, as well as the Galactic clumps from \citet[][solid black circles]{Wu2010} and the sample of nearby galaxies presented in \citet[][open black squares]{Gao+Solomon2004}. The red box represents the estimate for the Milky Way's $L_{\rm HCN}$ and $L_{\rm IR}$ derived using the \higal catalogue, while the paler box represents the estimate after accounting for emission from diffuse gas (see Section\,\ref{HCN from diffuse gas} for details). Small squares show extra-galactic GMCs, same as in Figure\,\ref{fig:Wu_slopes}. Dotted lines mark $L_{\rm IR}=10^{4.5}\,{\rm L}_\odot$ and the estimates of the Milky Way's total infrared luminosity \citep{Cox1986} and total molecular gas mass \citep{Heyer&Dame2015}.}
    \label{fig:MW_estimate}
\end{figure}

\higal provides mass measurements for $119\,923$ clumps, yielding a combined clump mass of $\log_{10}(M_{\rm dust})=7.84\,\text{M}_\odot$. We multiply this value by a factor of 1.25 to account for the \simm30\,000 clumps in the \higal catalogue that do not have mass measurements available. This provides an estimate for the total dense-gas mass of the Galaxy of $\log_{10}(M_{\rm dust})=7.94\,\text{M}_\odot$. Using the relationship between the HCN luminosity and dense-gas mass determined in Section\,\ref{HCN as a mass tracer}, we estimate the total contribution of Hi-GAL clumps to the HCN luminosity to be $\log_{10}(L_{\rm HCN}) = 6.4\pm 0.28\,{\rm K\,km\,s^{-1}\,pc^2}$. We note that this estimate does not include a correction for the mass-completeness limit of the \higal survey, which increases with heliocentric distance \citep{Elia2017} and therefore results in undetected low-mass clumps at large distances not being included in the catalogue. Our derived Galactic dense-gas mass should therefore be considered a lower limit. We further note that while \higal covers the full 360\degr\ range of Galactic longitude, its latitude coverage is restricted to a narrow strip around the Galactic mid-plane ($|b|\lesssim$1--2.5\degr\ following the Galactic warp; \citealt{Elia2021}), and any dense clumps located outside this latitude range are similarly not included in our estimate.

Similarly, we can obtain an estimate for the total contribution of the embedded young stars in the \higal catalogue to the infrared luminosity of the Galaxy. To do this, we only include protostellar clumps as defined by \higal (this would include \hii\,regions, YSOs, and protostellar clumps following the ATLASGAL classification), since these have significant 70-\micr emission and so actively contribute to the infrared luminosity. There are $\sim 35\,000$ protostellar clumps of which $\sim 5\,000$ do not have their distance constrained. The 30\,000 clumps that do have bolometric luminosity provide a total $\log_{10}(L_{\rm bol})=8.14\,{\rm L_\odot}$. Using the conversion between bolometric and IR luminosity derived in Section\,\ref{IRvBOL} we estimate their total $\log_{10}(L_{\rm IR})\approx8.49\,{\rm L_\odot}$. We increase this value by 15\,per\,cent to account for the other $\sim 5\,000$ protostellar clumps that do not have distances; this gives an estimate for contribution to the total infrared luminosity from embedded star formation of $\log_{10}(L_{\rm IR})\approx8.56\pm0.39\,{\rm L_\odot}$. As with the dense-gas mass estimate above, this value is subject to the same \higal completeness limit \citep{Elia2017} and restricted latitude coverage \citep{Elia2021}, and should therefore likewise be considered a lower limit on the true contribution of embedded star formation to the Galaxy's infrared luminosity.

In Figure\,\ref{fig:MW_estimate}, we show the IR and HCN luminosities of our full sample as well as those of \citet{Gao+Solomon2004} and \citet{Wu2005}. The total estimated contribution to the Milky Way's IR and HCN luminosities from \higal clumps calculated in the previous two paragraphs is also shown as a red box, the area of which represents the associated uncertainty in these measurements. It is clear from this plot that the IR and HCN luminosities of the dense gas not only do not satisfy the linear relationship determined by \citet{Wu2010} and \citet{Gao+Solomon2004}, but also fall short of the luminosity values reported for nearby galaxies by \citet{Gao+Solomon2004}.

Our measurement of the HCN luminosity is an order of magnitude lower and our IR luminosity is two orders of magnitude lower than other nearby galaxies. This is unexpected, given that previous estimates for the total IR luminosity (\citealt{Cox1986}; $\sim 1.5\times10^{10}$\,L$_\odot$) and total molecular mass ($\sim1\times10^9\,{\rm M_\odot}$ independently estimated by \citealt{Heyer&Dame2015} and \citealt{Csengeri2016_plank}, corresponding to $\log_{10}(L_{\rm HCN}) \approx 7.6\,{\rm K\,km\,s^{-1}\,pc^2}$) are both in line with measurements of other spiral galaxies in the \citet{Gao+Solomon2004} sample (these two values are shown on Figure\,\ref{fig:MW_estimate}).

In the following sections, we discuss why our approximation of these values differs so drastically from the other Milky Way estimates and other galaxies, and therefore what the contribution of dense clumps to the overall luminosity of the Galaxy is.

\subsection{Where is the missing HCN luminosity?} \label{HCN from diffuse gas}

In the previous section, we present an estimate of the contribution of dense clumps to the overall HCN luminosity of the Milky Way. Our value of $\log_{10}(L_{\rm HCN}) = 6.4\pm 0.28\,{\rm K\,km\,s^{-1}\,pc^2}$ is a factor of 2-3 smaller than the least massive galaxy in the \citet{Gao+Solomon2004} sample and a full order of magnitude lower than the estimate provided by \citet{Heyer&Dame2015}, however, their molecular gas mass is estimated from CO observations and therefore also likely to include a significant contribution from more diffuse gas. A direct comparison of our molecular mass estimate from dense gas with that estimated by \citet{Heyer&Dame2015} indicates that the dense gas fraction in the Milky Way is $\sim$10\,per\,cent of the total molecular gas. This is consistent with the conclusions of other studies of the dense gas fraction of large samples of Galactic molecular clouds \citep[e.g.,][]{Eden2013, Battisti2014}. If we accept that the dense gas fraction for the Milky Way is $\sim$10\,per\,cent and that the HCN luminosity is only tracing the dense gas, then either the nearby spiral galaxies in the \citet{Gao+Solomon2004} sample have dense gas fractions of order unity or, if they have similar dense gas fractions to the Milky Way, are an order of magnitude more massive. Both of these options seem unrealistic and it would seem more likely that the HCN luminosity measured for external galaxies includes a significant contribution from diffuse gas that is sub-thermally excited.

\citet{Stephens2016} also estimated the total HCN luminosity from dense gas and obtained a similarly low value. They propose that subthermal emission from gas below the HCN critical density could dominate the total $L_{\rm HCN}$ of galaxies. While they do not provide a robust estimate for the contribution from diffuse gas, their conclusion is supported by more recent observational and simulation-based studies \citep[e.g.,][]{Kauffmann2017, Pety2017, Barnes2020, Evans2020, Jones2023, Priestley2026}.

\citet{Kauffmann2017} use maps of the Orion A molecular cloud to investigate seven commonly observed molecular lines around 100\,GHz, including HCN. They compare the visual extinction across the region with the emission of molecular lines to find the characteristic (median) column density for each molecular line. They find that only half of the HCN emission comes from $A_{\rm v}$ greater than 6.1 magnitudes, which corresponds to volume densities of $n \sim 1000$\pcm. A follow-up survey of the far more extreme W49 molecular cloud, following the same method, finds a higher characteristic visual extinction of $A_{\rm v} \sim 12$\,mag, corresponding to volume densities of $3.4\pm2.8\times 10^3$\pcm \citep{Barnes2020}. While the variation in these values likely reflects the large difference in distance between the clouds and therefore different surface-brightness sensitivity, both works demonstrate characteristic densities much lower than the critical density of HCN, indicating that large-scale HCN observations do not exclusively trace dense gas. This supports the hypothesis that sub-thermal emission is a significant contributor to the total HCN luminosities of these nearby spiral galaxies.

\begin{figure}
    \centering
    \includegraphics[width=0.45\textwidth]{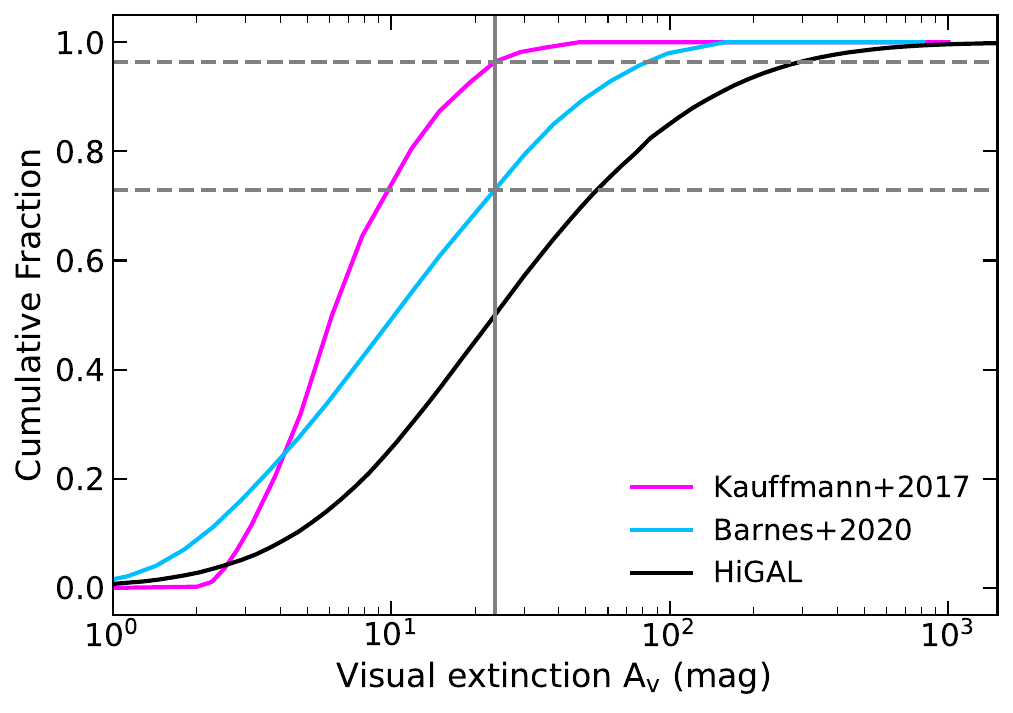}
    \caption{A cumulative distribution of the visual extinction of the \higal compact source catalogue as well as the Orion A \citep[][]{Kauffmann2017} and W49 \citep[][]{Barnes2020}. The characteristic extinction of \higal clumps (\simm 23.6\,mag) is marked with a vertical grey line. }
    \label{fig:visual extinction}
\end{figure}

To quantify the effect of their findings on our estimate of the total $L_{\rm HCN}$ of the Milky Way, we apply the analysis from \citet{Kauffmann2017} to our sample. We find that the median extinction of \higal\, clumps is $A_{\rm v} \sim 23.6$ (see Figure\,\ref{fig:visual extinction}). Comparing this to the characteristic visual extinction of the Orion A  and W49 molecular clouds of 6.1\,mag \citep{Kauffmann2017} and 12\,mag \citep{Barnes2020}, respectively, shows that our estimate represents about $5-25$ \,per\,cent of the total HCN emission of the Galaxy. This means that the Milky Way estimate derived previously can be increased by a factor of $\sim4-20$, resulting in $\log_{10}(L_{\rm HCN}) \approx 7.0 - 7.7\,{\rm K\,km\,s^{-1}\,pc^2}$. This is in good agreement with the total Galactic molecular gas mass estimated by \citet{Heyer&Dame2015} and places our Galaxy at a similar HCN luminosity to those of the nearby spiral galaxies studied by \citet{Gao+Solomon2004} (see the pale red box in Figure\,\ref{fig:MW_estimate}).  This leads us to conclude that extragalactic observations of HCN are also dominated by sub-thermal emission. While HCN is still broadly associated with denser regions than traced by $^{12}$CO, on galaxy scale observations, the emission is dominated by sub-thermal emission and is a poorer tracer (compared to N$_2$H$^+$) of the very dense gas that participates in star formation (\citealt{Kauffmann2017}). It has been shown, however, that at kpc scales HCN and N$_2$H$^+$ intensities trace each other \citep{Jimenez-Donaire2023}, implying a constant dense gas fraction on these scales. Therefore, while HCN itself is not a good dense gas tracer, it can be used to infer the amount of dense gas by accounting for the dense gas fraction.

\subsection{Where is the missing infrared luminosity?}
\label{IR Luminosity from Star-forming Clumps}

\citet{Cox1986} estimate that the Milky Way has a total $L_{\rm IR}\approx 1.5\times10^{10}$\,L$_\odot$. This value is consistent with the IR luminosity estimated for the nearby galaxies, placing it towards the lower end of the luminosity range. They estimate that \simm 37\,per\,cent of the total IR luminosity is associated with cold dust ($14-25$\,K) heated by the interstellar radiation field; \simm 50\,per\,cent originates from heating by O and B type stars, with the majority of the emission coming from extended low-density \hii regions; the remaining \simm13\,per\,cent is associated with hot dust ($250-500$\,K). More recently, \citet{Paladini2007} demonstrate that \simm80\,per\,cent of the far-infrared emission in the Milky Way is associated with atomic gas, further highlighting the dominant role of the diffuse ISM in the Galactic infrared output.

The IR luminosity we have estimated from the embedded star formation population associated with \higal clumps falls far short of this value ($L_{\rm IR} \approx3.5\times 10^8\,{\rm L_\odot}$), corresponding to only $\sim$2.4\,per\,cent of the  Galactic IR luminosity estimated by \citet{Cox1986}. This is consistent with the findings of \citet{Elia2025}, who show that compact sources account for only $\sim$2\,per\,cent of the total emission in 70\,\micr\ Hi-GAL maps. We have shown previously, in Section\,\ref{Gao-Solomon Relation}, that the IR luminosity is dominated by \hii regions and that these clumps do broadly satisfy the \citet{Wu2010} and \citet{Gao+Solomon2004} star formation relation. Within our sample, \hii regions account for 96\,per\,cent of the total IR luminosity but only 74\,per\,cent of the total HCN luminosity, despite making up just 45\,per\,cent of clumps, indicating that \hii region clumps contribute disproportionately more to the total IR luminosity than to the total HCN luminosity. However, there are only $\sim$1250 embedded \mbox{\hii\ regions} in the Galaxy (\citealt{Urquhart2022}) and so these represent $<1$\,per\,cent of all clumps in the Galaxy.

Our sample of 126 \hii regions (\simm 45\,per\,cent of our complete sample) has a mean $\log_{10}(L_{\rm IR})=5.24\,{\rm L_\odot}$. Assuming this mean IR luminosity of our sample of \hii regions, we estimate the total contribution from the embedded \hii region population of roughly 1250 is $\log_{10}(L_{\rm IR})\approx8.34\,{\rm L_\odot}$, which is the majority of the total we have determined for the whole \higal clump population. This is consistent with embedded \mbox{\hii\ regions} contributing the vast majority of the infrared luminosity of \higal clumps. It also means that embedded \mbox{\hii\ regions} contribute only  \simm1.5\,per\,cent to the estimated total IR luminosity of the Galaxy. This is consistent with the results of \citet{Cox1986}, where they consider compact \hii regions to contribute insignificantly to the overall luminosity of the Galaxy. In contrast, assuming the mean HCN luminosity of our \hii-region sample ($\log_{10}(L_{\rm HCN})=1.65\,{\rm K\,km\,s^{-1}\,pc^2}$), the same 1250 \hii\ regions would contribute only $\sim$0.14\,per\,cent to the estimated total HCN luminosity of the Galaxy ($\log_{10}(L_{\rm HCN})\approx{7.6}\,{\rm K\,km\,s^{-1}\,pc^2}$; see Section\,\ref{estimating total luminosities}).

The estimate of the total IR luminosity from \higal clumps clearly demonstrates that the overall IR luminosity of galaxies does not come from the embedded star formation taking place in dense clumps and is therefore not associated with ongoing star formation. The remaining IR flux may result from larger-scale structures such as classical \hii regions and their associated PDRs, which are not actively forming stars, but are the remnants of previous star formation activity. If that is indeed the case, then a significant portion of the emission would originate from more evolved O and B type stars and would be proportional to their lifespan \citep[$\sim$3-5\,Myr;][]{Woosley2002}. Since \hii\,regions only spend $\sim110\,000\,$yr \citep{Sabatini2021} in their embedded phase our estimate can be increased by a factor of $\sim20$ to get the total contribution to the IR luminosity from OB type stars throughout their life. Therefore, half of the IR luminosity of our Galaxy is associated with massive stars, which is consistent with the conclusions of \citet{Cox1986}. This means that extra-galactic observations trace much longer dynamical times (\simm4\,Myr) than clump-scale observations ($\ll\,1$\,Myr; \citealt{Krumholz2012}). It follows that the strong correlation demonstrated by \citet{Gao+Solomon2004} can therefore be interpreted as a result of constant star formation efficiency over long timescales and large spatial scales.

\subsection{Reinterpreting the IR-HCN relation}

The results presented in this work suggest that the commonly assumed interpretation of the IR-HCN correlation as a direct link between dense gas and ongoing star formation requires revision. While a linear relationship between $L_{\rm IR}$ and $L_{\rm HCN}$ has been reported on galaxy and GMC scales \citep[see ][ for a compilation of observations to date]{Neumann2025}, the correlation breaks down when the full population of Galactic clumps is considered. Although clumps that fully sample the initial mass function (IMF) broadly reproduce the relationship reported in previous studies, inclusion of the wider clump population yields a significantly steeper relation, demonstrating that the observed scaling is sensitive to sample selection and evolutionary state.

The estimates of the Milky Way HCN and infrared luminosities provide insight into the origin of this discrepancy. If the extragalactic IR-HCN relation directly traced the connection between dense gas and embedded star formation, integrating the contribution of all dense clumps in the Galaxy should reproduce the luminosities observed in nearby spiral galaxies. Instead, the luminosities derived from the \higal\ clump population fall substantially below both the values expected from the \citet{Gao+Solomon2004} relation and independent estimates of the total Galactic luminosity and molecular gas content.

The difference in HCN luminosity can be reconciled by accounting for emission from lower-density gas. Studies of Galactic molecular clouds indicate that a significant fraction of HCN emission originates from gas with densities well below the critical density of the transition, implying substantial sub-thermal excitation \citep{Kauffmann2017, Barnes2020}.This suggests that galaxy-scale HCN observations are dominated not solely by dense star-forming material but also by more extended, moderately dense gas.

The shortfall in infrared luminosity points to a similar conclusion. Embedded star formation within dense clumps contributes only a small fraction of the total infrared luminosity of the Galaxy. Instead, most of the IR emission likely arises from extended \hii\ regions, PDRs, and dust heated by the broader stellar population. Consequently, extragalactic infrared luminosities trace star formation integrated over several Myr rather than only the embedded phase associated with dense clumps.

Taken together, these results imply that the IR-HCN relation is not fundamentally a direct connection between dense gas and instantaneous star formation activity. Rather, it reflects the relationship between the total reservoir of moderately dense molecular gas and star formation averaged over large spatial and temporal scales. The approximate linearity of the relation may therefore arise because both the dense gas fraction and star formation efficiency remain relatively constant when averaged over whole galaxies and GMC complexes. Although HCN itself does not directly trace the dense gas participating in star formation, observations indicate a relatively constant dense gas fraction on kpc scales \citep{Jimenez-Donaire2023}, allowing HCN to act as an indirect proxy for the amount of dense gas. Similarly, the physical origin of the infrared luminosity changes with spatial scale: on clump scales it is associated primarily with embedded star formation, whereas on galactic scales it is dominated by emission from more evolved stellar populations over timescales of several Myr. The persistence of the IR-HCN relation therefore also requires a relatively constant star formation rate over these timescales. As a result, the correlation remains valid when averaged over large scales, but breaks down on clump scales where larger-scale components are filtered out and local variations are no longer averaged over.

\section{Summary and Conclusions}\label{Conclusion}

We present new HCN observations towards 426 dense clumps identified in the ATLASGAL survey to investigate the linear relationship that connects dense gas and star formation in galaxies, GMCs and dense clumps (e.g., \citealt{Gao+Solomon2004,Wu2005, Chen2017}).  Previous studies of Galactic clumps have focused on infrared bright clumps, leading to these samples being dominated by \hii regions and it is not clear if this relation will hold for a more representative sample of clumps. We combine our results with a similar set of observations of clumps located in the fourth quadrant by \citet{Stephens2016}, resulting in a sample of 592 clumps that cover the full range of evolutionary stages (quiescent clumps to those hosting \hii regions). After excluding complex profiles (blended and self-absorbed spectra), this is reduced to 343 clumps, of which 280 have been classified into one of four evolutionary stages (i.e., quiescent, protostellar, YSO, and \hii regions).

A summary of our key findings is as follows:

\begin{itemize}

    \item We find strong correlations between both the IRAS infrared and bolometric luminosities ($r_{s}\sim$0.8), and between HCN luminosity and clump mass ($r_{s}\sim$0.9). The strong correlation between the IR and bolometric luminosities allows us to extend the study to lower luminosities and earlier evolutionary stages. Using a linear fit we determined the  mass to HCN ratio to have a value of \simm30, which is 3 times higher than commonly used in extragalactic studies, suggesting that they include emission from gas below the critical density of HCN.\\

    \item Fitting the infrared and HCN luminosities for our whole sample of clumps returns a slope of $1.80\pm0.08$, which is significantly steeper than the linear fits reported in the literature. Breaking this into evolutionary sub-samples we find shallower slopes ($\sim$1.2), however, these are still steeper than values in the literature. Although previous studies have found a linear relationship, these have tended to focus on the most luminous clumps, while our analysis has shown this is not consistent with a more representative sample of Galactic clumps.\\

    \item Using a fully representative Galactic-wide sample of dense clumps we have found that the total infrared and HCN luminosities both fall significantly short of expected values. Therefore, extra-galactic observations cannot be explained through dense clumps alone, and other more dominant contributors to the luminosities must be accounted for.\\

    \item The total HCN luminosity of \higal clumps is \simm5 times lower than the least luminous of the other late-type spiral galaxies. Taking into account HCN emission from low and moderate density gas increases the total HCN luminosity by a factor of $\sim$5, closing the gap between the Milky Way and similar spiral galaxies. However, this indicates that the vast majority of gas traced by HCN in nearby galaxies is coming from sub-thermal emission from lower density gas, which is not directly associated with current star formation.\\

    \item The total infrared luminosity from embedded star formation, determined from the \higal clumps, only accounts for about 3 per\,cent of the total expected Galactic infrared luminosity, with the vast majority being produced by evolved \hii regions and PDRs. This indicates that the majority of infrared emission in galaxy studies is not linked to ongoing star formation, but instead traces longer timescales.\\

    \item The extragalactic studies of GMC-size scales and larger all have results consistent with a linear relation. However, we have shown that the HCN luminosity is dominated by emission from lower density gas, much of which is emitting sub-thermally, and the infrared luminosity is dominated by evolved \hii\ regions that have long since dispersed their natal clouds. The observed linear relation in nearby galaxies therefore reflects a nearly constant star formation efficiency averaged over large areas and long timescales, rather than a direct link between dense gas and SFR.\\

\end{itemize}

\section*{Acknowledgements}
The authors kindly thank the anonymous reviewer for their thoughtful comments, which have improved the quality and clarity of this work. We also thank Ian Stephens, Ashley Barnes, and Aashini Patel for helpful discussions and insights that contributed to this work. This work was supported by the Science and Technology Facilities Council (STFC) [grant number ST/X508469/1]. W.-J.\,Kim, F.\,Wyrowski, and D.\,Colombo acknowledge support by the Deutsche Forschungsgemeinschaft (DFG) via the Collaborative Research Center (CRC) SFB 1601 ``Habitats of massive stars across cosmic time'' (subproject B1). A.\,Karska acknowledges support from the Polish National Science Center SONATA BIS grant No. 2024/54/E/ST9/00314.

\appendix

\section*{Data Availability}
Full versions of Table 1 and spectral plots of the data are available from CDS at \url{http://cdsweb.u-stras bg.fr/}.




\bibliographystyle{mnras}
\bibliography{references} 





\bsp	
\label{lastpage}

\end{document}